\documentclass[aps,rmp,reprint,notitlepage,nofootinbi,floatfix,usenames,11pt.twocolumn,superscriptaddress]{revtex4-1}

\usepackage[T1]{fontenc} 
\usepackage{ae,aecompl}
\usepackage{caption}
\usepackage{graphicx}
\usepackage{placeins}
\usepackage{gensymb}
\usepackage{amsmath}
\usepackage[all]{nowidow}
\usepackage[colorlinks=true,allcolors=blue]{hyperref}

\def\apj{ApJ}%
\def\mnras{MNRAS}%

\usepackage{array}

\newcommand{\sv}{\langle \sigma v \rangle}

\usepackage{tikz}
\def\checkmark{\tikz\fill[scale=0.4](0,.35) -- (.25,0) -- (1,.7) -- (.25,.15) -- cycle;} 
\usepackage{multirow}
\usepackage{caption}

\begin{document}

\title{Dark Matter Annihilation to Neutrinos}


\author{Carlos A. Arg{\"u}elles}
\email{carguelles@fas.harvard.edu}
\affiliation{Department of Physics \& Laboratory for Particle Physics and Cosmology, Harvard University, Cambridge, MA 02138, USA\looseness=-10}

\author{Alejandro Diaz}
\email{diaza@mit.edu}
\affiliation{Department of Physics, Massachusetts Institute of Technology, Cambridge, MA 02139 USA\looseness=-10}

\author{Ali Kheirandish}
\email{kheirandish@psu.edu}
\affiliation{Department of Physics \& Center for Multimessenger Astrophysics\, Institute for Gravitation and the Cosmos, The Pennsylvania State University, University Park PA 16802 USA\looseness=-1}

\author{Andr\'es Olivares-Del-Campo}
\email{andres181192@gmail.com}
\affiliation{Institute for Particle Physics Phenomenology (IPPP), Durham University, Durham, UK\looseness=-10}

\author{Ibrahim Safa}
\email{isafa@fas.harvard.edu}
\affiliation{Department of Physics \& Wisconsin IceCube Particle Astrophysics Center, University of Wisconsin, Madison, WI 53706, USA\looseness=-10}
\affiliation{Department of Physics \& Laboratory for Particle Physics and Cosmology, Harvard University, Cambridge, MA 02138, USA\looseness=-10}

\author{Aaron C. Vincent}
\email{aaron.vincent@queensu.ca}
\affiliation{Department of Physics, Engineering Physics and Astronomy, Queen's University, Kingston, ON K7L 3N6, Canada\looseness=-10}
\affiliation{Arthur B. McDonald Canadian Astroparticle Physics Research Institute, Kingston, ON K7L 3N6, Canada\looseness=-10}
\affiliation{Perimeter Institute for Theoretical Physics, Waterloo, ON N2L 2Y5, Canada\looseness=-10}

\begin{abstract}
We review the annihilation of dark matter into neutrinos over a range of dark matter masses from MeV$/c^2$ to ZeV$/c^2$. Thermally-produced models of dark matter are expected to self-annihilate to standard model products.
As no such signal has yet been detected, we turn to neutrino detectors to constrain the ``most invisible channel.''
We review the experimental techniques that are used to detect neutrinos, and revisit the expected contributions to the neutrino flux at current and upcoming neutrino experiments.
We place updated constraints on the dark matter self-annhilation cross section to neutrinos $\sv$ using the most recently available data, and forecast the sensitivity of upcoming experiments such as Hyper-Kamiokande, DUNE, and IceCube Gen-2.
Where possible, limits and projections are scaled to a single set of dark matter halo parameters for consistent comparison.
We consider Galactic and extragalactic signals of $s$, $p$, and $d$-wave annihilation processes directly into neutrino pairs, yielding constraints that range from $\sv \sim 2.5\times10^{-26}~{\rm cm}^3 {\rm s}^{-1}$ at 30 MeV$/c^2$ to $10^{-17}~{\rm cm}^3{\rm s}^{-1}$ at 10$^{11}$ GeV$/c^2$.
Experiments that report directional and energy information of their events provide much stronger constraints, outlining the importance of making such data public.
\end{abstract}


\newcommand{\refeq}[1]{Eq.~(\ref{#1})}
\newcommand{\refeqs}[2]{Eqs.~(\ref{#1})~and~(\ref{#2})}
\newcommand{\refeqss}[3]{Eqs.~(\ref{#1}), (\ref{#2})~and~(\ref{#3})}
\newcommand{\reffig}[1]{Fig.~\ref{#1}}
\newcommand{\reffigs}[2]{Figs.~\ref{#1}~and~\ref{#2}}
\newcommand{\refsec}[1]{Section~\ref{#1}}
\newcommand{\refsecs}[5]{Sections~\ref{#1},~\ref{#2},~\ref{#3},~\ref{#4},~and~\ref{#5}}
\newcommand{\refappsec}[1]{Appendix Section~\ref{#1}}
\newcommand{\refapp}[1]{Appendix~\ref{#1}}
\newcommand{\reftab}[1]{Table~\ref{#1}}
\newcommand{\refref}[1]{~\cite{#1}}
\newcommand{\refrefs}[2]{~\cite{#1}~and~\cite{#2}}

\maketitle
\tableofcontents

\section{Introduction\label{sec:intro}}

The Standard Model (SM) of particle physics is the framework that describes matter and its interactions at the most fundamental level.
Despite overwhelming success as a predictive theory, observations indicate that the SM is incomplete.
Neutrinos have nonzero masses, yet the Higgs mechanism that provides masses for the other SM fermions cannot account for the chiral nature of neutrinos and their interactions unless additional particle content is added to the model.
Additionally, overwhelming astrophysical and cosmological evidence points to the existence of a new species of weakly-interacting particles -- dark matter (DM) -- which accounts for $\sim 85\%$ of the mass budget of the Universe.
Local stellar dynamics, galactic rotation curves~\cite{1970ApJ...159..379R,Persic:1995ru}, cluster dynamics~\cite{1937ApJ....86..217Z,1936ApJ....83...23S}, and gravitational lensing~\cite[\textit{e.g.}][]{Jee:2008qj,Jee:2007nx} all point to mass-to-light ratios in astrophysical objects that are much higher than could be accounted for by stellar objects and gas~\cite[for a historical overview see][]{Bertone:2016nfn}.
Measured primordial abundances of light elements tell us that Big Bang Nucleosynthesis requires a total baryon density\footnote{By \textit{baryonic} we refer here to stable nonrelativistic matter made of SM particles including neutrons, protons and electrons.} of only $\Omega_b \sim 0.05$, while the Cosmic Microwave Background (CMB) and other probes of large scale structure require the \textit{total} density of nonrelativistic matter to be $\Omega_m \sim 0.3$.
\footnote{More precisely, the baryon density is inferred to be $\Omega_b h^2 = 0.0224 \pm 0.0001$ and the (cold) DM density is $\Omega_c h^2 = 0.120 \pm 0.001$~\cite{Aghanim:2018eyx}, where $\Omega_i$ is the ratio of the density of component $i$ to the critical density, and the Hubble constant is $H_0 \equiv h$100 km s$^{-1}$Mpc$^{-1}$.} 

A leading hypothesis for the nature of this new nonbaryonic component is the Weakly Interacting Massive Particle (WIMP).
The relic abundance of WIMPs today was set as they fell out of equilibrium with the high-temperature plasma of the early Universe.
When the temperature, $T$, fell below the DM mass, $m_\chi$\footnote{We work in natural units where $c = \hbar = k_B = 1$.}, the equilibrium distribution became Botlzmann-suppressed, namely $\sim \exp(-m_\chi/T)$.
At some point, the expansion rate $H(t)$, became larger than the thermally-averaged self-annihilation rate, preventing further annihilation into SM particles, \textit{freezing-out} the relative density of DM particles.
The WIMP scenario predicts the observed relic abundance of DM for values of the thermally-averaged self-annihilation rate $\sv \simeq 3\times 10^{-26}$ cm$^{3}$s$^{-1}$ regardless of the final annihilation channel. 

Thermal production of weakly-interacting DM in the Early Universe implies possible ongoing self-annihilation to SM particles wherever DM exists today.
Significant effort has gone into searches for indirect signatures of DM annihilation.
Annihilation to most SM states yields an abundance of photons with energies on the order of $10\%$ of the DM mass, such that some of the strongest constraints on particle DM models are from the (non) observation of X- and gamma-ray signals from the Milky Way and its satellite galaxies; see \textit{e.g.}~\cite{Fermi-LAT:2016uux,Hoof:2018hyn}.
Cosmic-ray signatures provide similarly constraining limits, reports of excesses notwithstanding; see~\cite{Boudaud:2019efq} and references therein.

As X- and gamma-ray experiments rely, by design, on electromagnetic signals, they are optimal for probing links between the dark sector and quarks or charged leptons, although neutrino detectors can still play a role in these searches~\cite{Cappiello:2019qsw}.
There is a distinct possibility, however, that the principal portal through which the DM interacts with the SM is via the neutrino sector~\cite{Blennow:2019fhy}. 
This naturally arises in ``scotogenic'' models, in which neutrino mass generation occurs through interactions with the dark sector~\cite{Boehm:2006mi,Farzan:2012sa,Escudero:2016tzx,Escudero:2016ksa,Hagedorn:2018spx,Alvey:2019jzx,Patel:2019zky, Baumholzer19}.
These models introduce heavy neutrino states, sometimes called dark neutrinos, which could also provide a possible explanation of the MiniBooNE anomaly~\cite{Bertuzzo:2018itn,Ballett:2018ynz,Ballett:2019cqp,Ballett:2019pyw}. 
``Secret'' neutrino interactions with dark matter have recently become a very active field of investigation, where constraints have been obtained using high-energy astrophysical neutrinos~\cite{Farzan:2014gza,Davis:2015rza,Cherry:2016jol,Arguelles:2017atb,Kelly:2018tyg,Farzan:2018pnk,Pandey:2018wvh,Choi:2019ixb,Capozzi:2018bps,Murase:2019xqi}, solar neutrinos~\cite{Capozzi:2017auw}, cosmology~\cite{Campo:2017nwh,Barenboim:2019tux}, accelerator neutrino experiments~\cite{Aguilar-Arevalo:2017mqx,Arguelles:2018mtc,Hostert:2019iia} and colliders
~\cite{Primulando:2017kxf}.

Neutrinos are light, neutral, and notoriously difficult to detect.
If DM annihilates to heavy states such as muons, quarks, or weak bosons, a neutrino signal will be produced.
Unless annihilation occurs in an optically thick environment, the associated photon signal will always be easier to detect.
We thus choose to focus on the most invisible channel: direct annihilation of DM into neutrino-antineutrino pairs, whose energy will be equal to the DM rest mass, {\it i.e.} $E_\nu = m_\chi$.

The past two decades have seen extraordinary progress in the field of neutrino physics.
Observations span a wide energy range, from the MeV $pp$ solar neutrino flux~\cite{Agostini:2018uly} to the PeV ($10^6$ GeV) high-energy astrophysical neutrinos~\cite{Aartsen:2013jdh,Aartsen:2014gkd,Schneider:2019ayi}.
Furthermore, limits exist all the way up to $\sim$ ZeV ($10^{12}$ GeV) ~\cite{Aab:2015kma,Aartsen:2018vtx}.
With these observations, a multitude of experimental constraints have been derived on the DM annihilation cross section to neutrino pairs, either by experimental collaborations themselves or by independent authors recasting results of previous searches.
The goal of this work is to collect, when available, existing constraints on the $\chi \chi \rightarrow \nu \bar \nu$ annihilation channel, and otherwise to compute such limits from available data.
We focus on the two most promising sources of DM annihilation signal: 1) the dark matter halo of the Milky Way, in which we are deeply embedded, and 2) the full cosmic flux from the sum of all DM halos within our cosmological horizon.

Our main results are a set of constraints on a constant ($s$-wave) thermally averaged annihilation cross section $\sv$.
Where possible, we also compute constraints on $p$-wave ($\sv \propto (v/c)^2$) and $d$-wave ($\sv \propto (v/c)^4$) suppressed annihilations. These results are provided in Figures~\ref{fig:Indirect}--\ref{fig:Indirect_dwave}.
We cover a mass range from 1~MeV to $10^{15}$~MeV.
While the upper limit is a function of experimental reach, neutrino-coupled dark matter is severely constrained below $\sim 10$ MeV based on its modification of $N_{eff}$, the energy density in relativistic particles during nucleosynthesis~\cite{Kolb:1986nf, Serpico:2004nm, Boehm:2012gr,Ho:2012ug,Steigman:2013yua,Boehm:2013jpa,Nollett:2013pwa,Nollett:2014lwa, Steigman:2014pfa,Wilkinson:2016gsy,Escudero:2018mvt,Sabti:2019mhn}. 

The neutrino flux from DM annihilation depends sensitively on the DM halo shape, and many different assumptions have been employed, some in contradiction with kinematic observations~\cite{Benito:2019ngh}.
We thus embark on the endeavour to rescale or recompute all constraints using a single set of DM halo parameters.
Depending on the nature of the study and the available data, this is not always possible; when this is the case we explicitly mention it.
We provide, in the final section, estimates on the uncertainties associated with the choice of DM halo parameters.

This work contains the most up-to-date constraints.
While a few experiments come close in certain narrow mass ranges, it remains clear that current observations are not yet able to probe annihilation cross sections that explain the observed relic abundance of DM through thermal freeze-out.
This leaves plenty of room open for future searches, which is why we also present a forecast of possible limits from upcoming neutrino experiments~\cite{Arguelles:2019xgp}.

The structure of this review is as follows: we begin in Sec.~\ref{sec:theory} with a review of the annihilation signal we are constraining, from the Milky Way halo in Sec.~\ref{sec:galactic} and from the isotropic background of extragalactic halos in Sec.~\ref{sec:extragalactic}.
In Sec.~\ref{sec:pwave}, we detail the calculations needed to extend our analysis to velocity-dependent annihilations, namely $p$-wave and $d$-wave processes.
Sec.~\ref{sec:methods} briefly summarizes the experimental techniques used for neutrino detection in a wide energy range, and describes the statistical methods employed in this work to constrain the neutrino flux from dark matter annihilation.
Our results are presented in Sec.~\ref{sec:results}, including results from previous analyses that we recast to be consistent with our halo assumptions, wherever possible.
Sec.~\ref{sec:haloparams} shows the results of varying these assumptions in the range allowed by stellar dynamic observations for the Galactic component and simulation results for the extragalactic one.
Finally, we conclude in Sec.~\ref{sec:conclusion}.

\section{Dark matter annihilation\label{sec:theory}}

Neutrinos are the most weakly interacting stable particles in the SM and, consequently, the hardest to detect.
In the context of indirect detection, this implies that models where DM annihilates predominantly to neutrinos are difficult to rule out.
This makes the study of neutrinos as a final state particle particularly interesting as, so far, all direct and indirect searches for the footprints of DM--SM interactions have come up empty~\cite{Arcadi:2017kky,Tanabashi:2018oca}.
The limits derived on the DM annihilation to neutrinos can be interpreted as an upper bound on the total DM annihilation cross section to SM particles~\cite{Beacom:2006tt,Yuksel:2007ac}, since the latter is larger by definition. 

From a particle physics point of view, the direct annihilation of DM to neutrinos at tree level requires the addition of a neutrino-DM term to the SM Lagrangian that couples them.
Since neutrinos belong to an $SU(2)$ doublet, na\"ive SM gauge invariance implies that coupling neutrinos with DM would also induce an interaction between the DM and the charged leptons, mediated, \textit{e.g.}, by a new $Z$-like particle.
Such interactions are highly constrained, as they lead to production of dijet or dilepton signatures observable at colliders (see \textit{e.g.}~\cite{Carena:2004xs,Lees:2014xha}), fixed target experiments~\cite{Abrahamyan:2011gv}, and direct detection experiments (see \textit{e.g.}~\cite{Blanco:2019hah} and references therein).

Nevertheless, there exist viable models in which the DM phenomenology is dominated by its interactions with neutrinos~\cite{Blennow:2019fhy}.
Coupling only to the heavier lepton generations can strongly mitigate bounds from electron interactions, \textit{e.g.} by introducing a $U(1)_{L_\mu - L_\tau}$ symmetry~\cite{He:1990pn,He:1991qd}.
A more elegant option allows the DM to interact with a sterile neutrino that then mixes with the active neutrinos, leading to direct annihilations of DM to neutrinos if the mass of the sterile neutrino is larger than the DM mass~\cite{Profumo:2017obk,Ballett:2019pyw}.
If the sterile-light mixing is sizable, DM--neutrino interactions will provide the best window to understand such DM models.
A comprehensive review of these scenarios can be found in~\cite{Blennow:2019fhy}.

Finally, we are considering direct annihilation to neutrinos without including electroweak (EW) corrections, which severely complicate the spectral shape computations. These are important at energies above the electroweak scale, and will have two main consequences: 1) the peak of the spectrum will be slightly broadened, and 2) A lower-energy continuum will be produced.
Given the typical energy resolution $\gtrsim 10\%$~\cite{Aartsen:2013vja}  for high-energy neutrino detectors, the former effect is not likely to be important.
The second effect could potentially lead to stronger bounds from the additional flux at  lower energies.
A detailed computation of this effect up to ultra-high-energies has only recently been performed~\cite{Bauer:2020jay}; as these were not available at the time of this analysis we do not include these here.
At sub-TeV energies, these corrections are accurately implemented in numerical codes such as PYTHIA~\cite{Sjostrand:2014zea,Sjostrand:2019zhc}; a comparison between our limits and ones derived using these additional corrections show very little difference~\citep[see e.g.][]{Liu:2020ckq}.

A more important consequence is the presence of gamma radiation from the decay of EW products, which can potentially provide complementary constraints to dedicated neutrino-line searches~\cite{Murase:2012xs}. 
Using these secondary products, current constraints on the thermally averaged annihilation cross section to neutrinos from Fermi-LAT and HESS hover around $10^{-23}~{\rm cm}^3{\rm s}^{-1}$ in the 300~GeV to 3~TeV mass range~\cite{Queiroz:2016zwd}.
These gamma-ray based constraints are at the same level as current bounds from ANTARES~\cite{Adrian-Martinez:2015wey}, but are expected to be improved by the next generation gamma-ray experiments such as the Cherenkov Telescope Array (CTA)~\cite{Queiroz:2016zwd}.
We will provide an example using these projections for CTA in Sec.~\ref{sec:results}, noting that this only includes prompt gamma rays. 
Inverse-Compton scattering of primary electrons and positrons  with interstellar photons will strengthen the sensitivity of gamma-ray searches.
This effect has been studied for DM decay searches, but not for annihilation $\chi \chi \rightarrow \nu \bar \nu$~\cite{Cohen:2016uyg, Murase:2015gea, Chianese:2019kyl}.

\subsection{Galactic contribution\label{sec:galactic}}

We begin by setting limits on DM annihilation to neutrino pairs in the Milky Way (MW) dark matter halo.
The expected flux per flavor of neutrinos plus antineutrinos at Earth, assuming equal flavor composition\footnote{If the flavor composition at the source is not democratic, neutrino oscillation will yield a flavor composition at Earth that is close, but not equal to $(\nu_e:\nu_\mu:\nu_\tau) = (1:1:1)$.
Annihilation to $\nu_e$ only will give $\sim (0.55: 0.25:0.2)$; to $\nu_\mu$: $\sim (0.25: 0.36:0.38)$ and $\nu_\tau$ yields $\sim (0.19: 0.38:0.43)$. 
}, is given by
\begin{equation}
    \frac{d\Phi_{\nu+ \bar{\nu}}}{dE_\nu} = \frac{1}{4 \pi}
    \frac{\sv}{\kappa m_\chi^2}  \frac{1}{3}  \frac{dN_\nu}{dE_\nu} J(\Omega),
    \label{eq:galaxyAnnRate}
\end{equation}
where $\kappa$ is 2 for Majorana DM and 4 for Dirac DM, $m_\chi$ is the DM mass, and $\sv$ is the thermally averaged self-annihilation cross section into all neutrino flavors.
Going forward we set $\kappa = 2$ (Majorana DM). 
The spectrum in the case of annihilation to two neutrinos is simply ${dN_\nu}/{dE_\nu} = 2\delta(1 - E/m_\chi)m_\chi/E^2$.
$J(\Omega)$ is a three-dimensional integral over the target solid angle in the sky, $d \Omega$, and the distance $dx$ along the line of sight (l.o.s.) of the DM density $\rho_\chi$, namely
\begin{equation}
     J \equiv  \int d\Omega \int_{\mathrm{l.o.s.}}  \rho_\chi^2(x) dx. 
 \label{eq:Jfactordef}
\end{equation}
It is referred to as the $J$-factor and has units of GeV$^2$\,cm$^{-5}$\,sr.\footnote{Another equivalent convention used in the literature is to report the dimensionless quantity $\mathcal{J} = J/\Delta\Omega R_0 \rho_0^2$~\cite{Yuksel:2007ac}.}

The galactocentric distance is 
\begin{equation}
    r = \sqrt{R^2_0 - 2 \, x \, R_0 \, \cos{\psi} + x^2},
\end{equation}
where $\psi$ is the angle between the Galactic center (GC) and the line of sight, and $R_0$ is the distance from the Sun to the GC.
In practice, the upper limit of integration can be set at
\begin{equation}
    x_{\rm max} = \sqrt{R^2_{\rm halo} - \sin^2{\psi} R^2_0}+R_{\rm 0}\cos{\psi},
\end{equation}
for some maximum halo radius $R_{\rm halo}$.
The $J$-factor remains approximately unchanged for $R_{\rm halo} \gtrsim 30$ kpc.

To parametrize the DM halo, we use a generalized Navarro-Frenk-White (NFW) profile, which is given by
\begin{equation}
    \rho_{\chi}(r) = \rho_s \frac{2^{3-\gamma}}{\left(\frac{r}{r_s}\right)^\gamma \left(1+ \frac{r}{r_s}\right)^{3-\gamma}}.
    \label{eq:NFW}
\end{equation}
We take the Sun to be located $R_0 = 8.127~{\rm kpc}$  from the GC, as determined by recent measurements of the four-telescope interferometric beam-combiner instrument GRAVITY~\cite{Abuter:2018drb}.
We use DM halo parameters compatible with the best-fit values of~\cite{Benito:2019ngh}, {\it i.e.}: a local density\footnote{It is customary to specify $\rho_0 \equiv \rho_\chi(R_0)$ rather than $\rho_s$, as the former can be more directly measured. The two are related by inverting Eq. \eqref{eq:NFW}.} of $\rho_0 = 0.4~{\rm GeV~cm}^{-3}$, a slope parameter $\gamma = 1.2$, and a density $\rho_s$ at  scale radius $r_s = 20~{\rm kpc}$.
The resulting $J$-factors for $s$, $p$, and $d$-wave annihilation are shown in Tbl.~\ref{tab:Jtable}; the latter cases will be discussed in Sec.~\ref{sec:pwave}.
Some experiments, such as ANITA, AUGER, and GRAND, are only sensitive to a certain region of the sky.
In these cases, the corresponding $J$-factors must be recomputed by converting their respective sensitivity from elevation/azimuth to galactic coordinates, and integrating over the resulting region. 
A value of the $J$-factor is not given for some experiments, where the flux cannot be factored out as in Eq.~\eqref{eq:galaxyAnnRate}. This could be due \textit{e.g}. to an energy-dependent acceptance.
These are also shown in Tbl.~\ref{tab:Jtable}. When the exposure is not a simple declination window, we provide the reference to where it can be obtained.
Recent works~\cite{Pato:2015dua, Benito:2016kyp, Karukes:2019jxv,Benito:2019ngh} have constrained the halo shape and density parameters, using observations of stellar dynamics in the MW.
In Sec.~\ref{sec:haloparams}, we illustrate the effect on the dark matter limits obtained in this work when varying these parameters within those constraints.\\

\begin{table*}[!ht]
    \begin{center}
    \makebox[\linewidth]{
    \begin{tabular}{c | c | c | c | c}
    \hline \hline 
         Experiment& Exposure & $J_s/{10^{23}}$ & $J_p/{10^{17}}$ & $J_d/{10^{11}}$  \\ \hline
        $\heartsuit$ All-sky & All-sky & $2.3$& $2.2$& $3.6$ \\
        $\heartsuit$ GRAND  & Fig. 24 of \cite{Alvarez-Muniz:2018bhp} & $0.28$ & $0.28$  & $0.46$ \\ 
        $\heartsuit$ ANITA & dec = $[1.5^{\circ} , 4^{\circ}]$ & $0.018$  & $0.018 $ & $0.028$ \\\hline 
        CTA \cite{Queiroz:2016zwd} & Galactic Center \cite{Queiroz:2016zwd} & $0.074$ & $0.12 $ & $0.16$  \\ \hline
        $\heartsuit$ TAMBO & Fig. 3 \& 4 of \citep{Romero-Wolf:2020pzh} & 0.0009 & $-$ & $-$\\ \hline
        $\heartsuit$ Auger & \begin{tabular}{c}${\rm zenith} = [90^{\circ}, 95^{\circ}]$\\${\rm zenith} = [75^{\circ},90^{\circ}]$ \\ ${\rm zenith} = [60^{\circ}, 75^{\circ}]$\end{tabular} & \begin{tabular}{c}0.10\\0.28\\0.27\end{tabular} & $-$ & $-$\\ \hline
        $\heartsuit$ P-ONE & \begin{tabular}{c}$\cos({\rm zenith}) = [-1, -0.5]$\\ $\cos({\rm zenith}) = [-0.5, 0.5]$\\ $\cos({\rm zenith}) = [0.5, 1]$\end{tabular} & \begin{tabular}{c}0.87\\1.2\\0.13\end{tabular} & \begin{tabular}{c}0.85\\ 1.2\\ 0.12\end{tabular} & \begin{tabular}{c}1.4\\ 2.0\\ 0.18\end{tabular}\\ \hline
        \hline
    \end{tabular}
    }
    \end{center}
    \caption{
    \textbf{\textit{J-factors for different experiments discussed in this work and their associated halo parameters.}}
    $J$-factors, given in units of GeV$^2$ cm$^{-5}$ sr, are computed according to Eq.~\eqref{eq:Jfactordef}.
    We use these to find the expected neutrino flux as described in Eq.~\eqref{eq:galaxyAnnRate}.
    Each row corresponds to a different experimental setup given its angular exposure.
    The first column names the experiment; the second column summarizes their angular acceptance; and the last three columns give the $s$-wave, $p$-wave, and $d$-wave $J$-factors, respectively.
    The hearts, $\heartsuit$, indicate new results given in this work.}
    \label{tab:Jtable}
\end{table*}

\subsection{Extragalactic contribution\label{sec:extragalactic}}

In addition to DM annihilation in the MW, annihilation of extragalactic dark matter integrated over all redshifts should provide a diffuse isotropic neutrino signal~\cite{Beacom:2006tt}.
As in the search for extragalactic background light, there are two contributions to this isotropic flux: 1) a ``background'' flux from the diffuse (non-collapsed) distribution of DM, whose rate grows with redshift as $ \Omega_{DM}^2 \sim (1+z)^6$,  and 2) a late-time contribution from the large overdensities in galactic halos. 

In this case, the expected flux of neutrinos plus antineutrinos per flavor at Earth from DM annihilation is given by
\begin{eqnarray}
\frac { d \Phi _ { \nu + \bar{\nu} } } { d E_{\nu} } &=& \frac {1 } {4 \pi  } \frac { \Omega _ { D M } ^ { 2 } \rho _ { c } ^ { 2 } \sv} { \kappa m _ { x } ^ { 2 } } \frac{1}{3}\nonumber \\ 
&& \int _ { 0 } ^ { z _ { u p } } d z \frac { \left(1 + G ( z )\right) (1+z)^3} { H ( z ) } \frac { d N _ { \nu + \bar{\nu} } \left( E ^ { \prime } \right) } { d E^ { \prime } },
\label{eq:extragalactic_flux}
\end{eqnarray}
where $H ( z ) = H_0\left[ ( 1 + z ) ^ { 3 } \Omega _ {m} + (1+z)^4\Omega_r + \Omega _ { \Lambda } \right] ^ { 1 / 2 } $ is the time-dependent Hubble parameter, $\rho _ {c}$ is the critical density of the Universe,  $\Omega_m$, $\Omega_r$, and $\Omega_\Lambda$ are respectively the fraction of $\rho_c$ made up of matter, radiation and dark energy.
While the upper limit on redshift, $z_{up}$, can in principle be as high as the neutrino decoupling time at $T \sim$~MeV, neutrinos produced at that epoch are redshifted to the point of being invisible to existing detectors.  
${ d N _ { \nu } \left( E ^ { \prime } \right) }/ { d E }$ is the neutrino spectrum at the detector, where $E ^ { \prime }$ ($E$) is the energy at the source (detector). 
The spectrum is related to the source production spectrum via a Jacobian transformation to take cosmological redshift into account, namely
\begin{eqnarray}
\frac { d N _ { \nu + \bar{\nu}} \left( E ^ { \prime } \right) } { d E^ { \prime }  } &=& 2 \frac{m _ { \chi }}{E^{' 2}} \delta \left( \frac{m _ { \chi }}{E ^ { \prime }} - 1 \right) \nonumber \\
&=& \frac { 2 } {  E } \delta \left[ z - \left( \frac { m _ { \chi } } { E } - 1 \right) \right].
\end{eqnarray}
In Eq.~\eqref{eq:extragalactic_flux}, $\sv $ is the thermally averaged cross section.
The first part of the factor $1+ G ( z )$  in the integrand of Eq.~\eqref{eq:extragalactic_flux} represents the isotropic background DM contribution, while $G(z)$ is the halo boost factor at redshift $z$.
It accounts for the enhancement to the annihilation rate in DM clusters and their evolution with redshift; and is given by
\begin{eqnarray}
    G(z) &=& \frac{1}{\Omega_{DM,0}^2\rho_c^2}\frac{1}{(1+z)^6}  \\ 
    &&\int dM \frac{dn (M,z)}{dM}\int dr 4\pi r^2 \rho_{\chi}^2(r).\nonumber
    \label{eq:overdensity} 
\end{eqnarray}
The first integral is over halo masses $M$ whose distribution is specified by the halo mass function (HMF), $dn/dM$, while the second integral is over the halo overdensities themselves.
We model the latter as self-similar NFW profiles whose densities and radii are specified by a concentration parameter uniquely determined by their mass and redshift.
The parametrization that we employ is based on fits to the MultiDark/BigBolshoi~\cite{2012MNRAS.423.3018P} simulations and can be found in Appendix B  of~\citet{Lopez-Honorez:2013lcm}.

Two uncertainties arise from the integral over $M$.
First is the choice of integration limits, specifically the lower limit, $M_{min}$.
This is because smaller halos are more concentrated, thus contributing more to the injected neutrino energy.
This  means that  choosing arbitrarily low-minimum halo masses results in unrealistic limits.
It is common in the literature to use $M_{min} = 10^{-6} M_\odot$ as a benchmark, although there is no data-driven motivation for this choice.
$M_{min}$ is not well-constrained, and will ultimately depend on model details~\cite{Shoemaker:2013tda,Cornell:2013rza}.
Therefore, in this work we pick $M_{min} = 10^{-3}M_\odot$ as a conservative limit choice.
In section~\ref{sec:haloparams}, we show the effect of varying $M_{min}$ down to $10^{-9} M_\odot$.
The other uncertainty arises from the choice of HMF, $dn/dM$, parametrization. We use the results of the N-body simulation by~\cite{2013MNRAS.433.1230W}, as parametrized in~\cite{Lopez-Honorez:2013lcm,Diamanti:2013bia}.
Several other HMF parametrizations are tested, and the uncertainties due to choice of HMF are quantified in Sec.~\ref{sec:haloparams}. 

\begin{figure*}[t!]
  
    \includegraphics[width=0.49\linewidth]{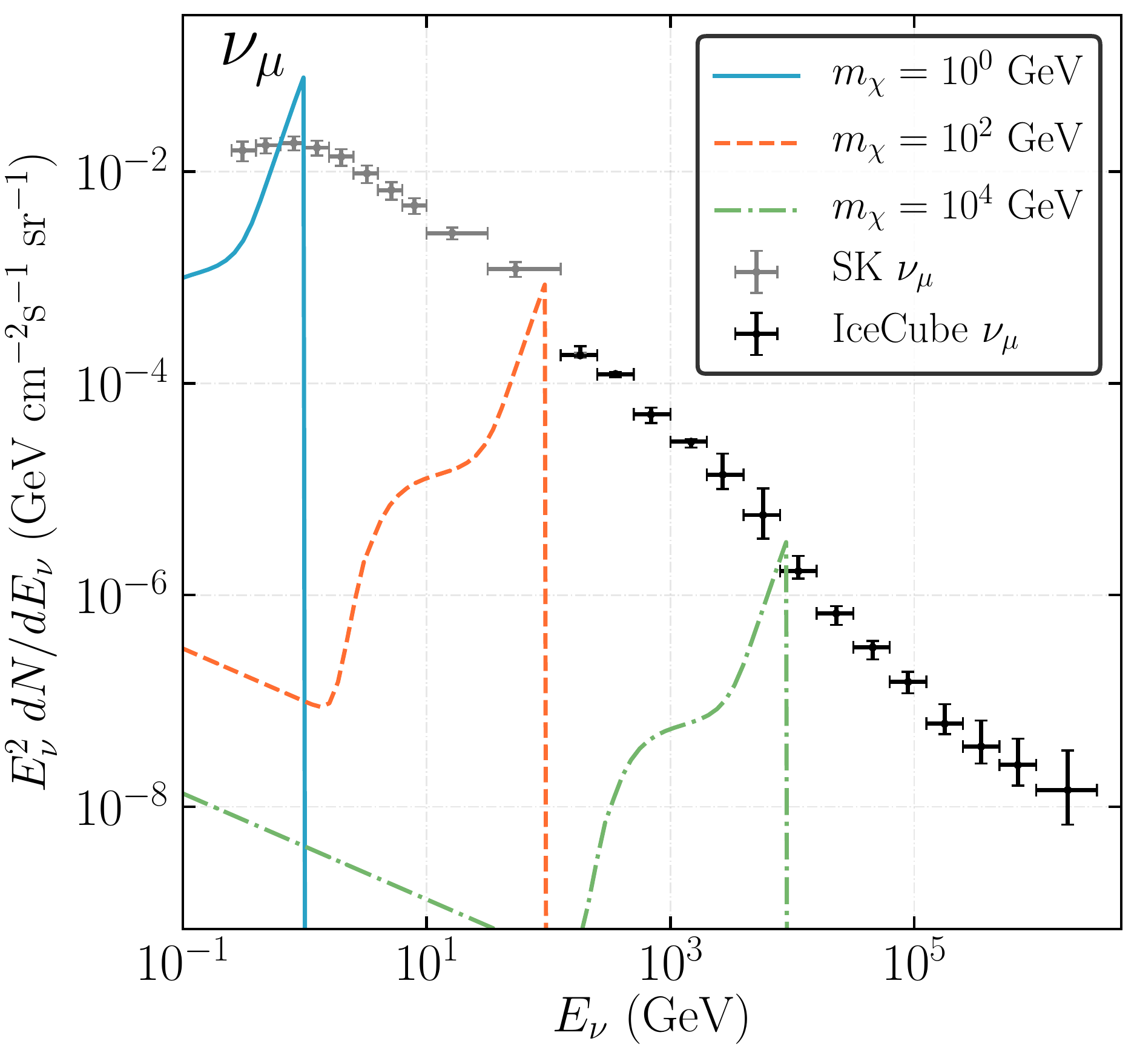}
    \includegraphics[width=0.49\linewidth]{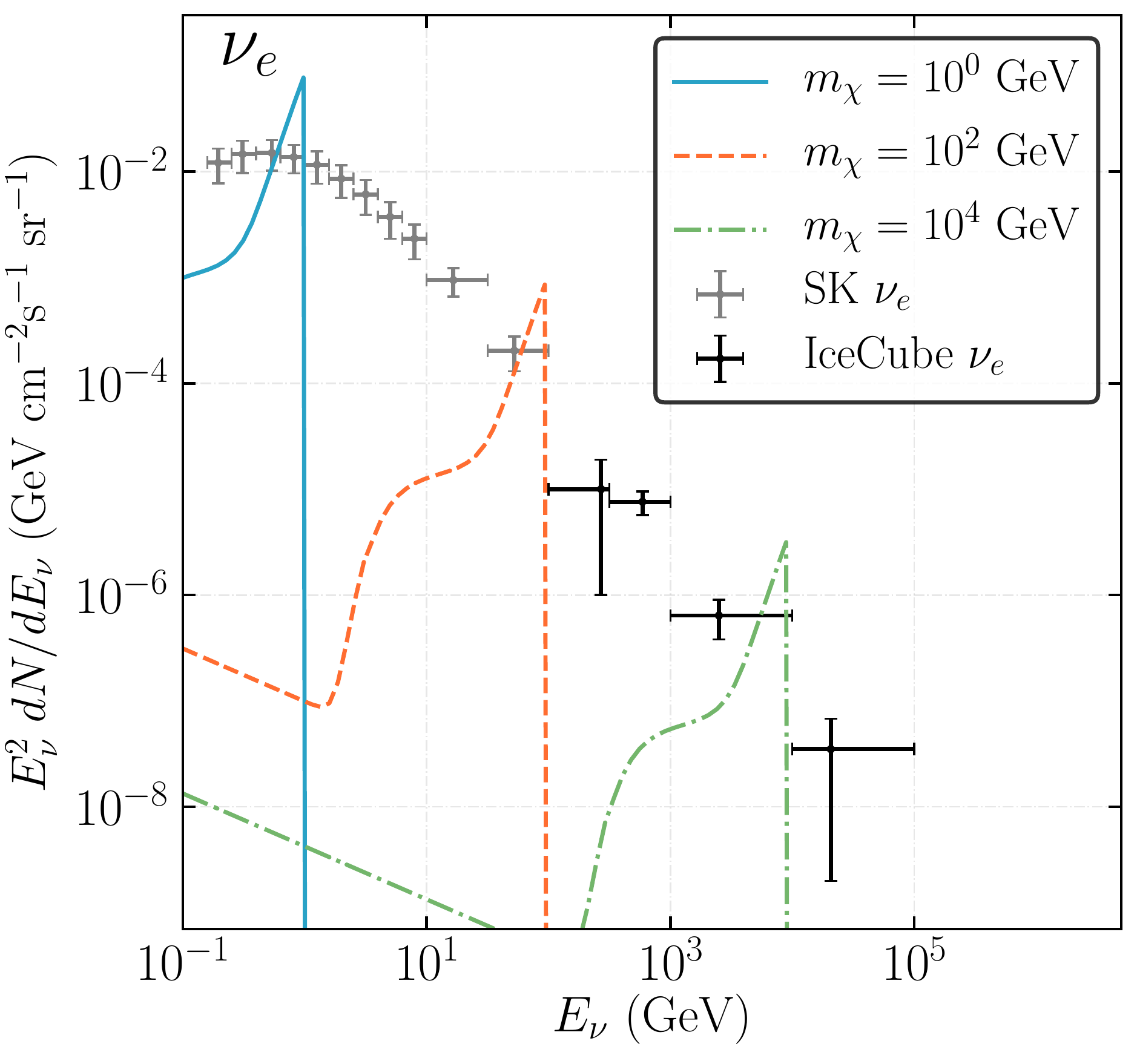}
    \begin{center}
    \includegraphics[width=0.51
    \linewidth]{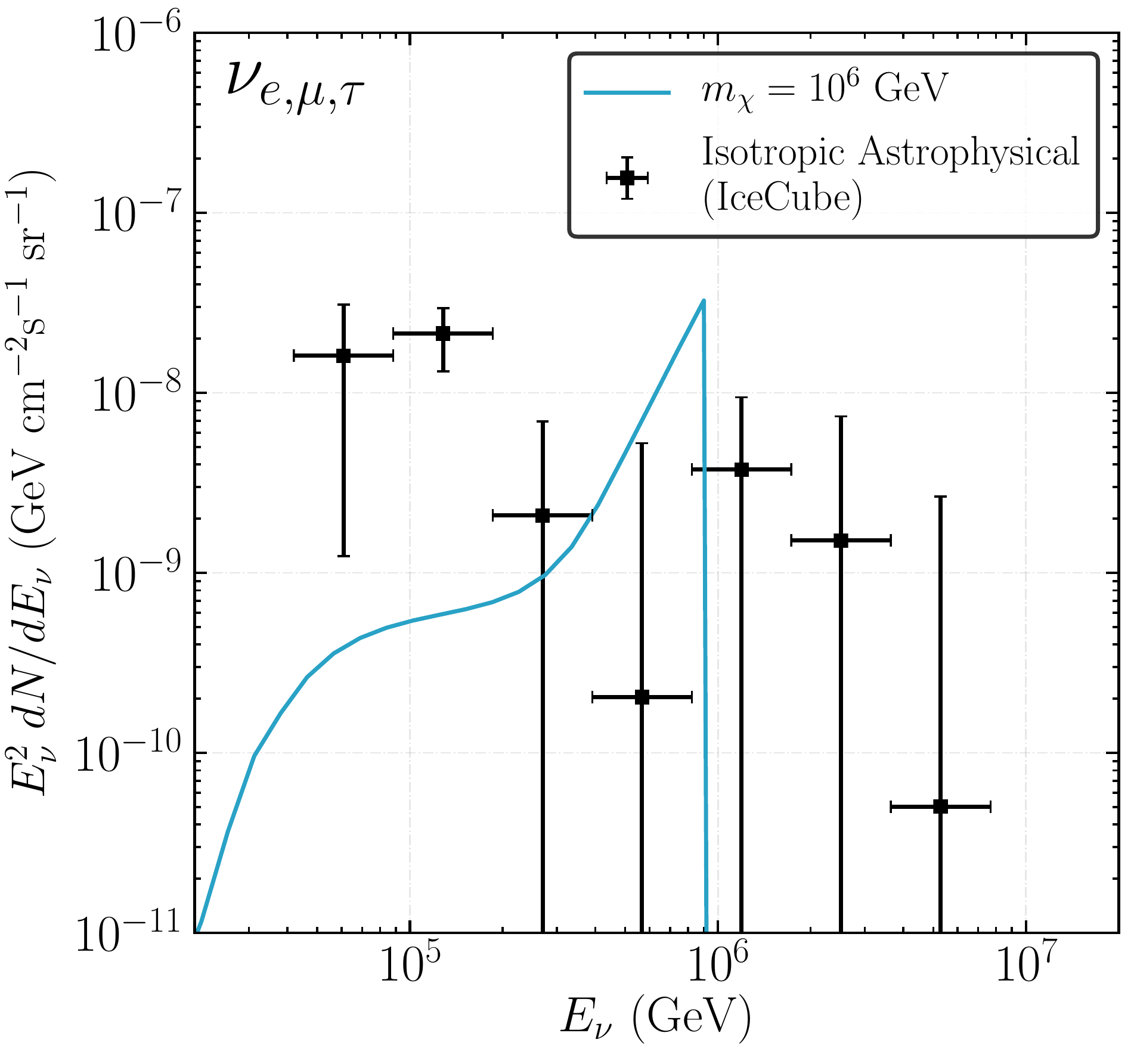}
    \end{center}
  \caption{
  \textbf{\textit{Examples of neutrino fluxes produced by dark matter annihilation overlayed on the observed neutrino distributions.}}
  Expected flux of neutrinos from extragalactic dark matter annihilation as a function of energy, shown for several dark matter masses.
  Fluxes are computed using the value of the cross section corresponding to the 90\% C.L. limit derived in this work.
  Here, the extragalactic dark matter annihilation fluxes are compared to the unfolded atmospheric fluxes from both Super-Kamiokande~\cite{Richard:2015aua} and IceCube~\cite{Aartsen:2015xup,Aartsen:2016xlq}.
  Top left is the $\nu_{\mu}$ channel; top right is the $\nu_{e}$ channel; the bottom shows a comparison to IceCube's measured per-flavor isotropic Astrophysical flux using 7.5 years of Starting Events~\cite{Abbasi:2020jmh}.
  }
  \label{fig:extragalacticflux}
\end{figure*}

The expected spectrum of DM annihilation to two neutrinos from cosmological sources is shown in Fig.~\ref{fig:extragalacticflux}, for different DM masses.
These are overlaid on the Super-Kamiokande (SK)~\cite{Richard:2015aua} and IceCube~\cite{Aartsen:2015xup,Aartsen:2016xlq} unfolded atmospheric $\nu_e$ and $\nu_\mu$ fluxes as well as the isotropic astrophysical flux~\cite{Abbasi:2020jmh}. 

\subsection{Velocity-dependent annihilation\label{sec:pwave}}

Certain matrix element vertex structures lead to a suppression of the constant ($s$-wave) part of the self-annihilation cross section.
Expanding in powers of $v/c$, the dominant term may be $p$-wave ($\propto v^2$) or $d$-wave ($\propto v^4$) in the nonrelativistic limit. 
The DM velocity distribution depends on the kinematic details of the structure in which it is bound, as well as its distance from the center of that distribution.
Assuming a normalized Maxwellian distribution, $f(v,r)$, with dispersion $v_0(r)$, the annihilation rate will be proportional to
\begin{equation}
\langle v^n \rangle = \int d^3v f(v,r) v^{n}.
\end{equation}
For $p$- and $d$-wave, this respectively yields 
\begin{eqnarray}
\langle v^2 \rangle &=& 3 v_0^2(r), \\
\langle v^4 \rangle &=& 15 v_0^4(r).
\end{eqnarray}
We obtain the dispersion velocity, $v_0$, by solving the spherical Jeans equation, assuming isotropy. This is given by
\begin{equation}
    \frac{d(\rho(r) v^2_0(r))}{dr}=-\rho(r) \frac{d\phi(r)}{dr} \, ,
    \label{eq:Jean}
\end{equation}
where $\phi(r)$ is the total gravitational potential at radius $r$.
For Galactic constraints, we include not only the contribution of the DM halo to $\phi(r)$, but also follow~\cite{Boddy:2018ike} and include a parametrization of the MW bulge and disk potentials to account for their masses.
These are given by
\begin{eqnarray}
\phi(r)_{\rm bulge}&=&-\frac{G_NM_b}{r+c_b}, \\
\phi(r)_{\rm disk}&=&-\frac{G_NM_d}{r} \left(1-e^{-r/c_d} \right),
\end{eqnarray}
where $G_N$ is Newton's gravitational constant, $M_b=1.5 \times 10^{10}M_{\odot}$, and $c_b =0.6$ kpc are the bulge mass and scale radius, while $M_d=7 \times 10^{10}M_{\odot}$ and $c_d=4$ kpc are the disk mass and scale radius~\cite{Boddy:2018ike}.
Galactic $J$-factors can then be reevaluated via
\begin{eqnarray}
J_{v^n} = \int d\Omega \int_{\mathrm{l.o.s.}}\frac{\left\langle v^{n}(r)\right\rangle}{c^{n}} \rho^2_\chi(r) dx.
\end{eqnarray}

In the case of our extragalactic analysis, we only include the potential from the DM halos themselves. This is conservative, in that the addition of the uncertain baryonic contributions would only strengthen our constraints.
In a similar manner to the Galactic case, Eqs.~\eqref{eq:extragalactic_flux} and \eqref{eq:overdensity} must be modified to include the dependence on $\langle v^n \rangle (r)$.
As long as the annihilation remains a two-to-two process \citep[unlike scenarios in e.g.][]{Bell:2017irk}, Eq.~\eqref{eq:extragalactic_flux} becomes:
\begin{eqnarray}
\frac { d \Phi _ { \nu } } { d E_{\nu} } &=& \frac {c } {4 \pi  } \frac { \Omega _ { D M } ^ { 2 } \rho _ { c } ^ { 2 } \sv} { 2 m _ { x } ^ { 2 } }  \\
&& \int _ { 0 } ^ { z _ { u p } } d z \frac { \left(\left[\frac{1+z}{1+z_{\rm KD}} \right]^{n} + G_n ( z )\right) (1+z)^3} { H ( z ) } \frac { d N _ { \nu } \left( E ^ { \prime } \right) } { d E }, \nonumber
\label{eq:extragalactic_flux_vn}
\end{eqnarray}
where the redshift $z_{\rm KD}$ is related to the temperature at kinetic decoupling $T_{\rm KD}$ and the temperature of the CMB today $T_{\rm CMB,0}$ via  $1 + z_{\rm KD} = T_{\rm KD}/T_{\rm CMB,0} \simeq 4.2 \times 10^9 \,  (T_{\rm KD}/\mathrm{MeV})$~\cite{Diamanti:2013bia}.
\citet{Shoemaker:2013tda} obtained a temperature of kinetic decoupling:
\begin{equation}
    T_{\rm KD} \simeq 2.02 \, \mathrm{MeV} \, \left(\frac{m_\chi}{\mathrm{GeV}} \right)^{3/4}.
    \label{eq:TKD}
\end{equation}
In general, kinetic decoupling occurs later than chemical freeze-out and depends on the number of relativistic degrees of freedom $g_\star(T_{\rm KD})$.
At redshifts where the annihilation products are still measurable by earth-based detectors, the factor of $((1+z)/(1+z_{\rm KD}))^n$ still leads to a strong enough suppression that it will always be subdominant to the halo contribution proportional to $G_n(z)$.
The exact value of $T_{\rm KD}$ in Eq.~\eqref{eq:TKD} is thus inconsequential. Eq.~\eqref{eq:overdensity} including velocity dependence is rewritten as follows:
\begin{eqnarray}
    G_n(z) &=& \frac{1}{\Omega_{DM,0}^2\rho_c^2}\frac{1}{(1+z)^6} \\ 
    && \int dM \frac{dn (M,z)}{dM}\int dr 4\pi r^2 \frac{\left\langle v^{n}(r)\right\rangle}{c^{n}} \rho_{\chi}^2(r), \nonumber
\label{eq:vdepboost}
\end{eqnarray}
where we have used the same HMF as in the velocity-independent case, with the addition of the velocity dispersion, $\left\langle v^{n}(r)\right\rangle$, in the rightmost integral.
~\cite{Diamanti:2013bia} provides the detailed method of solving the Jeans equation to compute $\left\langle v^{n}(r)\right\rangle$ as a function of the DM halo concentration.
For convenience, we provide the following function for the $p-$ and $d-$wave cases:
\begin{equation}
    \ln(G_n) \simeq \sum_{i} c_i~\alpha^i,
\label{eq:polyfit}
\end{equation}
where $c_i$ are the coefficients provided in Tbl.~\ref{tab:coeff}, and $\alpha \equiv \ln(z)$. This parametrization is valid down to redshifts  $\gtrsim 10^{-3}$.

\begin{table}[hb]
\centering
\begin{tabular}{ p{0.33cm} p{2cm} p{2cm} }
 \hline \hline
 & $p-$wave & $d-$wave \\
 \hline
 $c_0$ & $-7.004$   & $-19.88$\\
 $c_1$ & $-1.821$   & $-2.493$\\
 $c_2$ & $-0.5793$  & $-0.804$\\
 $c_3$ & $-0.09559$ & $-0.1636$\\
 $c_4$ & $-0.006148$  & $-0.02101$\\
 $c_5$ & $0$  & $-0.001181$ \\
 \hline \hline
\end{tabular}
\caption{\textbf{\textit{Coefficients of the polynomial fit to velocity dependent halo boost factors.}}
The coefficients corresponding to Eq.~\eqref{eq:polyfit}, which is a parametrization to the numerical solution of Eq.~\eqref{eq:vdepboost}.}
\label{tab:coeff}
\end{table}

\begin{figure*}[ht]
\centering
\includegraphics[width=\textwidth]{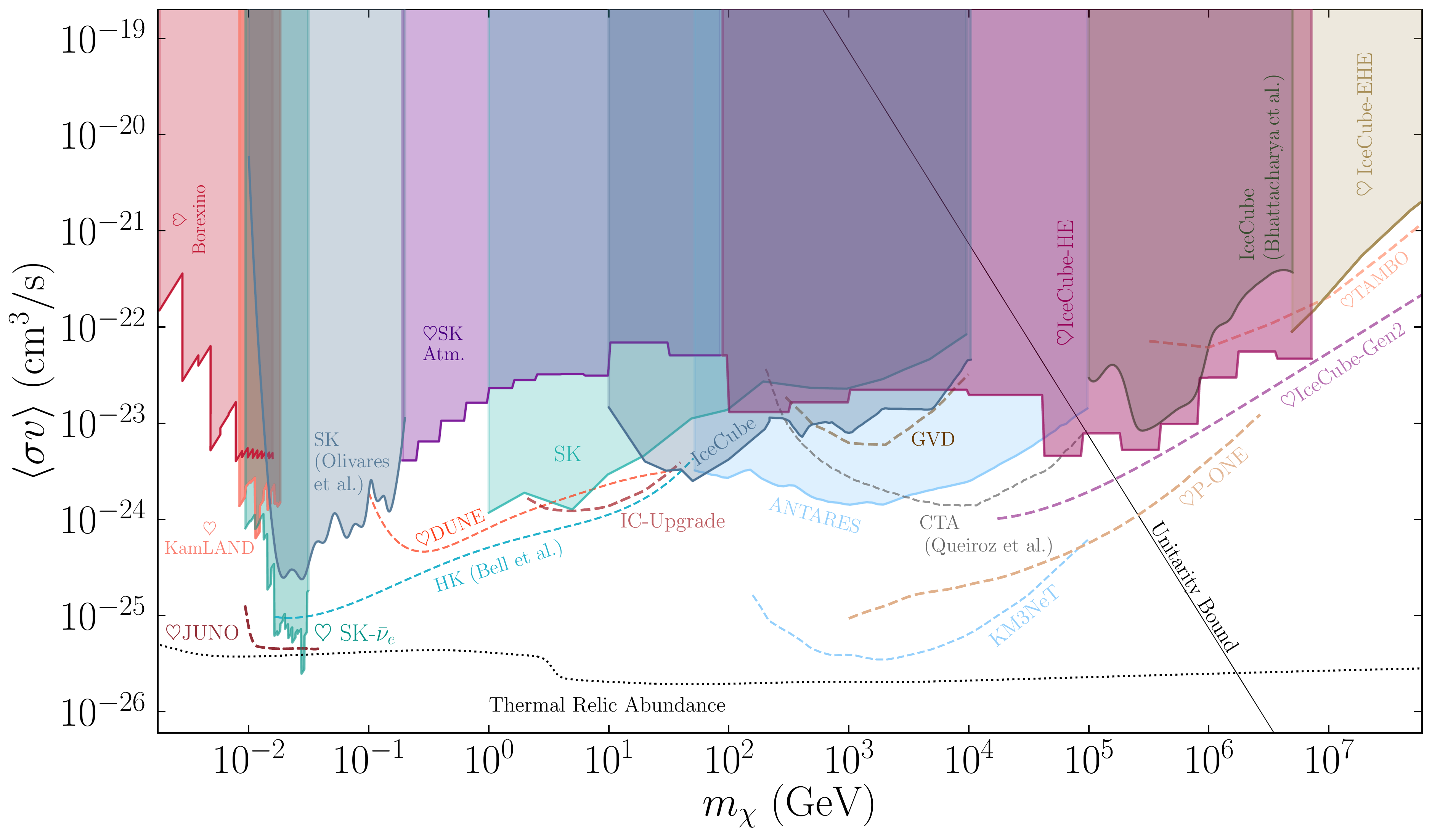}
\caption{\textbf{\textit{The landscape of dark matter annihilation into neutrinos up to $10^8$ GeV.}} We show results from this work, as well as previously published limits. Data and corresponding references are detailed in Sec.~\ref{sec:results}. Solid and dashed lines represent 90\% C.L. limits and sensitivities, respectively. Projected sensitivities assume five years of data taking for neutrino experiments and 100 hours of observation for CTA. The dotted line corresponds to the value required to explain the observed abundance via thermal freeze-out. The straight diagonal line, labeled as ``Unitarity Bound,'' gives the maximum allowed  cross section for a non-composite DM particle. These results assume $100\%$ of the dark matter is composed of a given Majorana particle. If instead only a fraction, $f$, is considered these results should be multiplied by $1/f^2$. In the case of Dirac DM, limits would be scaled up by a factor of two. The heart symbols ($\heartsuit$) indicate new results obtained in this work. See Fig. \ref{fig:Indirect_non_unitarity} for constraints and projections up to $10^{11}$ GeV.}
\label{fig:Indirect}
\end{figure*}

\section{Experimental Methods\label{sec:methods}}

In this section we will briefly review the different methodologies and technologies used for neutrino detection relevant for the discussion of the experimental results discussed in this review.
The results presented in Sec.~\ref{sec:results} rely on our understanding of the backgrounds in the region of interest.
Depending on whether the background flux is known, upper limits can be either background-agnostic or background informed.
Moreover, the upper limits highly depend on the systematics that govern neutrino detection, for instance the energy resolution and flavor identification capability.
Below, we first outline the statistical framework for limit-setting, before describing detector physics used over energy ranges considered here, from a few MeV up to $10^{12}$ GeV and beyond.

\subsection{Statistical Methods\label{sec:statistics}}

To contextualize the variety of experimental capabilities, we will first outline the principal statistical treatments used to infer the properties of the flux of neutrinos from dark matter annihilation.
We will explain them in increasing order of complexity and strength.

\subsubsection{Background-agnostic methods} In this method we use the observed data and the detector signal efficiency to constrain the flux of neutrinos from DM.
This method can inform us of the maximum allowed flux, but, by construction, it cannot be used to claim the observation of dark matter.
This technique is predicated on comparing the observed and expected number of events in a given bin, by means of the following likelihood function:
\begin{equation}
   \mathcal{L}(\mu) =  \left\{
                \begin{array}{l r}
                  \mathcal{P}(d | \mu) & (d < \mu),\\
                  1 & (d \geq \mu), \\
                \end{array}
              \right.
\end{equation}
for which the likelihood is less than one only if the predicted number of events $\mu$ is larger than the recorded data, $d$.
The probability distribution $\mathcal{P}$ could be a Poisson or Gaussian distribution depending on the sample size.
Using this likelihood one can construct one-sided confidence upper limits on $\mu$ and, in turn, on the dark matter cross section given the $J$-factor and detector acceptance.
The strength of this method is determined by experiment exposure, signal efficiency,  and the amplitude of unmodeled backgrounds; these determine the statistical uncertainty and the phase-space over which the bins are defined.
In the case of dark matter, one would ideally bin the events in: energy, direction, and morphology; but often this is either not done due to decreasing statistical power, insufficient Monte Carlo certainty, or increasing difficulty in modeling the systematic uncertainties.

In this review, we take advantage of this approach in a number of experimental settings.
As examples, we compare the Super-Kamiokande unfolded neutrino energy distribution~\cite{Richard:2015aua} to the dark matter expectation using this technique and perform a similar comparison to the IceCube PeV astrophysical neutrino segmented fit.
We also use this technique when experiments have not seen neutrino events and upper limits are reported, such as the Pierre Auger Observatory's limit on neutrino flux at very high energies.

\subsubsection{Background-informed methods}

A higher statistical power can be achieved by simultaneously modeling the signal -- the event rate due to dark matter -- and background -- any other contribution to the observed rate.
This requires signal and background efficiencies, as well as a model for the background distribution over each observable.
A prototypical likelihood function is:
\begin{equation}
    \mathcal{L}(\theta, \eta) = \mathcal{P}(d | \mu_s(\theta,\eta) + \mu_b(\eta)) \Pi(\eta),
\end{equation}
where $\mu_s(\theta,\eta)$ and $\mu_b(\eta)$ are the expected signal and background counts respectively, $d$ represents the observed counts, and $\theta$ and $\eta$ are the dark matter parameters and nuisance parameters, respectively. 
The latter parameters incorporate the effect of the systematic uncertainties in the signal and background distributions and are often constrained by previous knowledge or \textit{in situ} measurements represented in the function $\Pi(\eta)$.
When the signal and background predictions are well defined, the probability function, $\mathcal{P}$, is taken to be a Poisson function in the small-count regime or a Gaussian function in the large-count regime. 

If the signal or background predictions carry large uncertainties, which is often the case for rare backgrounds or signals that cover very specific parts of phase space such as dark matter lines~\cite{Gainer:2014bta}, stochastic likelihood models can be used~\cite{Glusenkamp:2017rlp,Arguelles:2019izp,Glusenkamp:2019uir}.
For other treatments proposed to tackle this problem see also~\citet{Barlow:1993dm,Bohm:2013gla,Chirkin:2013lya}.

In either case, the treatment of systematic uncertainties is often done by using the profile likelihood method, in which the likelihood function is maximized over the nuisance parameter at each physics parameter point~\cite{doi:10.1146/annurev.nucl.57.090506.123052}.
Alternatively, in Bayesian treatments~(see \textit{e.g}.~\citet{Trotta:2017wnx}) or hybrid frequentist-Bayesian treatments~\cite{Cousins:1991qz} the nuisance parameters are marginalized over by integrating the likelihood function.
In the case that the bin content is large, such that a Gaussian likelihood function is a good approximation, the expectations can be computed accurately.
Often, a multidimensional Gaussian is used where the covariance between bins incorporates both the systematic and statistical uncertainties.
The latter approach does not require additional parameters to incorporate systematic uncertainties into the likelihood, making it computationally advantageous.

With this formalism, background-informed analyses have additional power compared to the background agnostic scenario, provided that experiments are capable of constraining the background size, and separating it from signal.
The ability to constrain background is encapsulated in systematic uncertainties, whereas the separation of background from signal depends on the features of both.
The features in the case of neutrinos from dark matter are a democratic flavor composition, spatial clustering predominantly around the Galactic center, and an energy distribution which is maximal close to the dark matter mass.
Separating dark matter from background using these three features then depends on the experimental direction and energy resolutions, as well as its flavor identification capabilities dictated by the event morphological classification.
The latter is important since natural and anthropogenic sources often have a non-democratic flavor composition.
This is a characteristic of the stronger constraints.
For example, we use the fact that for MeV dark matter one of the main backgrounds are solar neutrinos, which can be efficiently removed by selecting only for antineutrinos in Super-Kamiokande or JUNO; we also rely on this in our predictions of the sensitivities for DUNE and Hyper-Kamiokande in the 100~MeV to 30~GeV energy range, where we use the fact that one can do morphological event analysis to remove muon neutrinos which are the dominant component of the atmospheric flux.

\subsection{Neutrino Detection Methods\label{sec:detectors}}

Because neutrinos only interact via the weak nuclear force, neutrino detection must proceed in at least two steps: first, interaction between a neutrino and a detector electron or nucleus, and second, the detection of the resulting electromagnetic signal. 
Typically, energy from a gamma-ray or electron cascades down via scintillation, additional ionization or Cherenkov radiation and is subsequently measured by optical sensors or charge readout.

The small neutrino detection cross section poses a great challenge in the search for the expected fluxes from dark matter annihilation to neutrinos.
As the dark matter mass increases, larger detectors are necessary to compensate for the smaller flux, which scales as $m_\chi^{-2}$.
Such a scaling can come at the cost of energy and angular resolution, as well as flavor identification, all of which allow differentiation between the dark matter induced neutrinos from other natural or anthropogenic neutrino sources as discussed in the previous section.
In this section, we review the techniques used to detect neutrinos in different energy ranges; see also~\cite{Katori:2016yel,Diaz:2019fwt} for a discussion in the context of neutrino oscillation experiments. Note that the energy ranges detailed here are approximate, and there is naturally some overlap between techniques and physics discussed in each respective subsection.

\subsubsection{Neutrino energies below 10~MeV}

Coherent elastic neutrino-nucleus scattering, namely $\nu A^Z_N \to \nu A^{*Z}_N $, dominates the cross section at the lowest energies \cite{Freedman:1973yd}.
This process, sometimes abbreviated as CE$\nu$NS, has no kinematic threshold and scales quadratically with the atomic number.
However, the maximum recoil energies are very small making its detection difficult; in fact it has only recently been observed using anthropogenic neutrinos in detectors of $\mathcal{O}(10)$~kg of mass~\cite{Akimov:2017ade}.
Future ton-scale dark matter direct detection experiments such as DARWIN~\cite{Aalbers:2016jon} expect to see solar and atmospheric neutrinos via CE$\nu$NS.
Because of the trade-off between detector size and nuclear recoil threshold, they would only be sensitive to DM above $m_\chi \sim 10$~MeV, and provide only marginal improvement over existing dedicated neutrino experiments that use different detection channels.

Neutrino-electron scattering also has no kinematic threshold at detectable energies, and the cross section is predicted without ambiguities that arise from form factors in hadron-neutrino interactions.
This interaction's well-understood kinematics, together with the fact that a single outgoing charged particle is produced, makes it a good channel to use for DM annihilation searches. 
This is because precise energy and directional information can be inferred. 
The angle between the neutrino and the electron is tightly constrained by the kinematics, $E_e \theta_e < 2 m_e$, allowing for an accurate reconstruction of the neutrino direction (it was through this process that in 1998 the Super-Kamiokande experiment made the first image of the Sun in neutrinos~\cite{Fukuda:1998fd}; see also~\citet{Ahmad:2001an,Alimonti:2000xc,Arpesella:2008mt} for subsequent measurements by SNO and Borexino).
Angular information is used to mitigate the $\sim$ 1-10 MeV solar neutrino backgrounds and to search for correlations with the expected angular distribution of DM via $J(\Omega)$.
Unfortunately, the neutrino-electron cross section is approximately $10^{-43}~{\rm cm^2}$ at 5~MeV, which is about a factor of 10 smaller than the dominant neutrino-nucleon process.

The other commonly-used technique to detect sub-10~MeV neutrinos is inverse beta decay (IBD), $\bar\nu_e p \to n e^+$.
This is due to three reasons: first, the large and well-measured IBD cross section, approximately $10^{-42}~{\rm cm^2}$ at 5~MeV~\cite{Vogel:1999zy,Ankowski:2016oyj}, with an uncertainty of $\sim 0.2\%$~\cite{Vogel:1999zy,Kurylov:2002vj}; second, the low-threshold: $E_\nu > 1.806~{\rm MeV}$; and finally, the ability to reduce background by searching for the prompt positron signature followed by the neutron capture.
This detection method is often used with hydrocarbon-based scintillator since it contains a large number of free protons and emits large number of photons, typically $10^4$ per MeV of deposited energy~\cite{leo1994techniques}.
The energy deposited by the prompt signal is the kinetic energy of the positron plus two 511~keV gamma-rays from electron-positron annihilation, and a 2.2~MeV gamma ray from the delayed capture of the neutron on free protons.
In hydrogen-based detectors the neutron capture time is typically 300~$\mu s$.
If the detector is doped with 1\% Gadolinium, this time is reduced to about 20~$\mu s$ and the prompt gamma-ray energy is 8~MeV allowing for an improved background suppression~\cite{Beacom:2003nk}; \textit{e.g.} in the case of Super-Kamiokande a hundredfold background suppression efficiency can be achieved~\cite{Watanabe:2008ru}.
In the search for dark matter this process has the advantage that it is only triggered by $\bar\nu_e$ allowing for very efficient suppression of the solar neutrino flux that dominates the natural backgrounds at sub-10 MeV energies.
In fact, our strongest limit across all dark matter masses comes from an IBD search by Super-Kamiokande; see Fig.~\ref{fig:Indirect}. 
 
\subsubsection{Neutrino energies between 10~MeV and 1~GeV}

Between $\sim 10$~MeV and $\sim 1$~GeV, in Cherenkov detectors the proton is invisible since it is Cherenkov threshold --  approximately 1.3~GeV in mineral oil, 1.4~GeV in water, and can be as low as 1.2~GeV in the Antarctic ice~\cite{2012TCD.....6.4695B}.
This has advantages and disadvantages compared to scintillator detectors, on the one hand it simplifies identification and classification of events since the observed Cherenkov light must be associated with the outgoing charged lepton.
On the other hand, the lack of proton kinematics means that the energy and angular resolution can be greatly degraded. 
The dominant neutrino-nucleon process in this energy range is that of charged-current quasi-elastic (CCQE) scattering, namely $\nu_\alpha N \to \alpha \tilde N$ where $\alpha$ is a charged lepton and $N$ ($\tilde N$) is a proton or neutron.
At high enough energies, muon neutrinos can have CCQE interactions, producing muons which can be identified by the morphology of the Cherenkov ring.
Due to the larger mass, muons tend to preserve their direction as they travel through the detector producing sharper rings than electrons.
Cherenkov detectors can be constructed out of mineral oil, water, or ice.
Although oil-based detectors boast a larger Cherenkov angle and the ability to run without a purification system, they are only utilized in smaller detectors~\cite{Diaz:2019fwt} due to the higher filling cost.
For this reason, multi-kiloton detectors available as of 2020 are all water or ice based. As early as 2022, JUNO will become the first multi-kiloton liquid scintillator detector. 
Since the DM-induced flux is expected to be very small, the larger water or ice Cherenkov detectors currently dominate the constraints over oil-Cherenkov detectors and we will not discuss them further.

\subsubsection{Neutrino energies from 1~GeV to $10^7$~GeV}

Resonant light-meson production is important between approximately 1 and 10~GeV.
Due to the difficulty in cross section modeling, neutrino detection in this range is subject to large uncertainties.
Above 10~GeV the contribution of deep inelastic scattering (DIS), where the neutrino exchanges a $W$ or $Z$ boson with one of the partons inside the nucleon becomes the dominant process.
The production of taus in tau-neutrino charged-current interactions becomes possible above the threshold $m_\tau = 1.777$~GeV, though the cross section is only around 15\% of the charged-current muon-neutrino cross section at 10~GeV, rising to 75~\% at 100~GeV~\cite{Conrad:2010mh}.

Though unsegmented Cherenkov detectors are still used in this energy range, the use of tracking calorimeters, often constructed as segmented scintillators, are popular as they allow for improved reconstruction of outgoing muon tracks, as well as electromagnetic and hadronic showers produced in the interaction vertex.
Notable examples of these types of detectors in contemporary neutrino physics are the NO$\nu$A experiment and the T2K near-detector. 
Sampling calorimeters have also been used to increase the target density, though this comes at the expense of a degraded energy resolution.
In this case a dense material like iron is interleaved with scintillator panels.
This design was used by the MINER$\nu$A experiment~\cite{Aliaga:2013uqz} to perform precision measurements of the neutrino cross section and has been used in the past to measure neutrino oscillations by MINOS~\cite{Sousa:2015bxa}.
In these detectors the morphological features observed in the trackers have been used to identify the different neutrino interaction processes by comparing them to generated event libraries~\cite{osti_925917,Backhouse:2015xva} or convolutional neural networks~\cite{Aurisano_2016,Psihas:2019ksa}.
Given the size of these detectors they are not expected to play a role in the detection of dark matter and are not included in this work.

The newest neutrino detectors in this energy range are the so-called liquid argon time projection chambers (LArTPC)~\cite{Cavanna:2018yfk}.
These detectors consist of an electric field cage filled with liquid argon.
When a charged particle is produced in the argon, it travels through the medium and ionizes the argon atoms, liberating electrons.
An electric field then drifts the electrons to wire planes on one side of the detector, recording a projected footprint of the interaction.
Three dimensional reconstruction is also possible by using the timing of the charge deposition on the wires. 
To localize the event in the third dimension, the drift time of electrons in argon and the initial interaction time need to be known.
The initial interaction time can be known in the case of generic neutrino interactions via the scintillation light produced by the charged particles in argon or, in the case of neutrinos produced in bunches in a beam, by the beam timing.
In the case of dark matter searches, relevant for this work, only the former technique is relevant.
Even though the neutrino-argon cross section is currently poorly understood compared to other  materials conventionally used in neutrino physics, these detectors have the potential for unprecedented particle identification: see \textit{e.g.}~\citet{Acciarri:2017hat,Adams:2018bvi,MicroBooNE:2018mfn}.
Examples of currently operating LArTPC neutrino detectors are MicroBooNE~\cite{Acciarri:2016smi} and ICARUS~\cite{Ali-Mohammadzadeh:2020fbd} at Fermilab.
The next generation experiment in this category is DUNE~\cite{Abi:2020evt}.

At the higher end of this energy range, neutrino telescopes such as ANTARES and IceCube have the largest neutrino collection volumes.
These detectors operate at energies above 10~GeV where DIS is the dominant cross section process~\cite{Gandhi:1995tf}.
These detectors use natural media, such as the Mediterranean water or the Antarctic ice, as targets for the neutrino interaction.
Cherenkov light produced by charged particles by products of these interactions are then observed by photomultiplier tubes (PMTs) arranged on sparse arrays. 
In these detectors the different neutrino interactions map onto different morphologies of the time and spatial distribution of charge in the array.
Neutral-current interactions, charged-current electron-neutrino interactions, and most of the charged-current tau-neutrino interactions produce a morphology known as a cascade.
Because cascades can be contained in the detector, this morphology has the best energy resolution.
Charged-current muon neutrino interactions produce a morphology known as tracks, due to the long travel-time of the muon.
This morphology provides the best directional information.
In water, photons tend to scatter less than in ice, providing more direct light. 
This means that the muon angular resolution in water-based detectors is better than those in ice.
On the other hand, given the longer absorption length of photons in ice compared to water, the effective detector volume is larger for detectors deployed deep in the ice.
Finally, charged-current tau neutrino interactions can produce a variety of morphologies depending on the boost factor of the tau and its decay channel.
For example, around 1~PeV, a tau can travel on average 50~m before decaying producing separated energy depositions known as {\em double bangs}~\cite{Learned:1994wg,Cowen:2007ny}; 
in 2018 IceCube announced the first candidate astrophysical tau events~\cite{stachurska_juliana_2018_1301122,Stachurska:2019wfb}.
Finally, in these detectors one can also observe the electron-neutrino scattering, since at approximately 6.3~PeV an electron antineutrino can resonantly scatter with an atomic electron producing a $W$ on shell~\cite{Glashow:1960zz,Loewy:2014zva}; 
$W$-production of coherent photon scattering can also be important at these energies see~\cite{Seckel:1997kk,Alikhanov:2015kla,Beacom:2019pzs,Zhou:2019vxt,Garcia:2020jwr}. 
The observation of this process provides a unique handle on the ratio of neutrinos to antineutrinos, as well as providing exquisite energy resolution; and in fact, a candidate event has recently been detected~\cite{2019ICRC...36..945L}.

\subsubsection{Neutrino energies above $10^7$~GeV}

At extremely-high energies, the neutrino flux expected from dark matter and other astrophysical sources such as cosmogenic neutrinos is very small, necessitating the construction of detectors with effective volumes much larger than a cubic kilometer.
Neutrino interactions in this energy range occur overwhelmingly via deep inelastic scattering~\cite{Gandhi:1995tf}.
Two main techniques are used to search for neutrinos in this energy range, both of which rely on identifying horizontal or upgoing particles to mitigate the larger cosmic-ray backgrounds.
The first method involves looking for air showers induced by neutrino-nucleus interactions in the atmosphere or just below the surface of the Earth, while the second uses the radio signature produced in very-high-energy neutrino interaction~\cite{Gusev_1984,Markov:1986dx}, known as Askaryan radiation~\cite{Askaryan:1962hbi,Zas:1991jv}.

This former technique can be detected in a number of ways: sparse surface arrays of water Cherenkov tanks are used to identify charged particles from showers as they develop over an area that may span many square-km.
Air fluorescence telescopes and optical air Cherenkov telescopes can also be used alone or in combination with water tanks~\cite[as is the case for Auger,][]{ThePierreAuger:2015rma}. 
The timing, morphology, and amount of light deposition is used to infer the energy of the incoming particle, its direction, and its nature.
In particular, a neutrino will typically travel much deeper into the atmosphere than a cosmic ray or gamma ray before interacting.
Tau neutrinos are particularly promising, as $\tau$ leptons can be produced in a nearby mountain or below the horizon~\cite{Jeong:2017mzv}.
If the tau survives the journey out of the mountain, its decay yields an upgoing air shower~\cite{Reno:2019jtr,Reno:2019qmk}; an EeV $\tau$ typical interaction length is a few kilometers in rock and is shorter than its decay length.
The expected event rate for such processes at cosmic ray observatories like Auger turns out to be higher than from neutrino-induced atmospheric showers, thanks to the high density of rock.
Radio arrays such as GRAND~\cite{Alvarez-Muniz:2018bhp} have been proposed to cover as large an effective area as possible (up to two-hundred thousand square-km) to search for such a signal.

The second method, Askaryan radiation detection, aims to observe neutrinos via the radio emission generated by charge displacement caused by the developing electromagnetic or hadronic shower after DIS scattering.
This emission is distinct from down-going cosmic-ray showers in that the polarization of the radio signal is expected to be different.
This technique has been implemented by using radio antennae either suspended from balloons~\cite{Gorham:2010kv} or buried in the ice~\cite{Anker:2020lre,Allison:2019xtn} in the Antartic continent.
The ability to cover a large area with a single antenna cluster makes this a very scalable and relatively low-cost technique. 

\section{Results\label{sec:results}}

\begin{figure*}[htp!]
\centerline{\includegraphics[width=\linewidth]{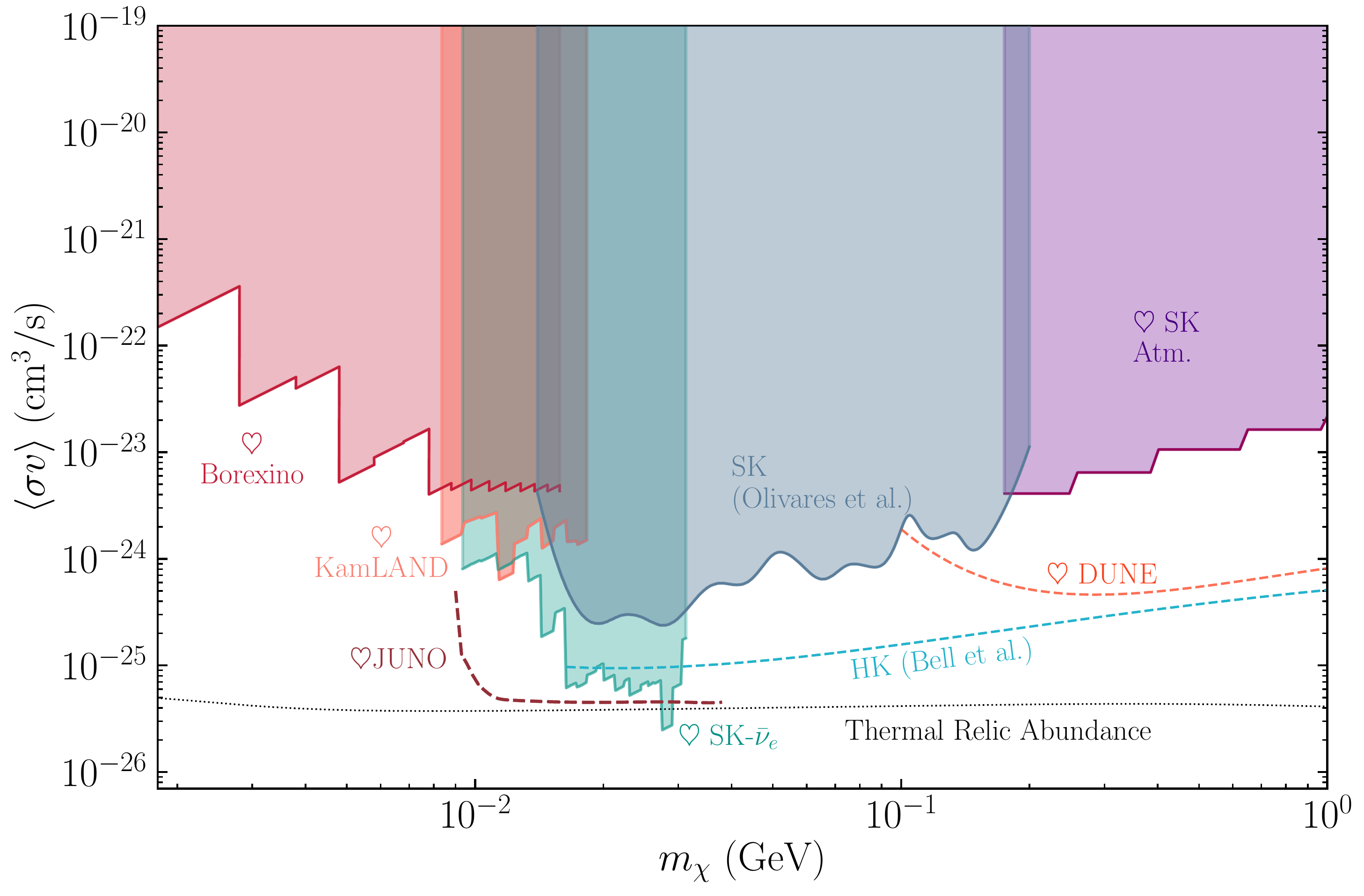}}
\caption{
\textbf{\textit{The landscape of sub-GeV dark matter annihilation into neutrinos.}}
Same as Fig.~\ref{fig:Indirect}, but restricted to dark matter masses below one GeV.
}
\label{fig:Indirect_lowe}
\end{figure*}

\begin{figure*}[ht!]
\centerline{\includegraphics[width=\linewidth]{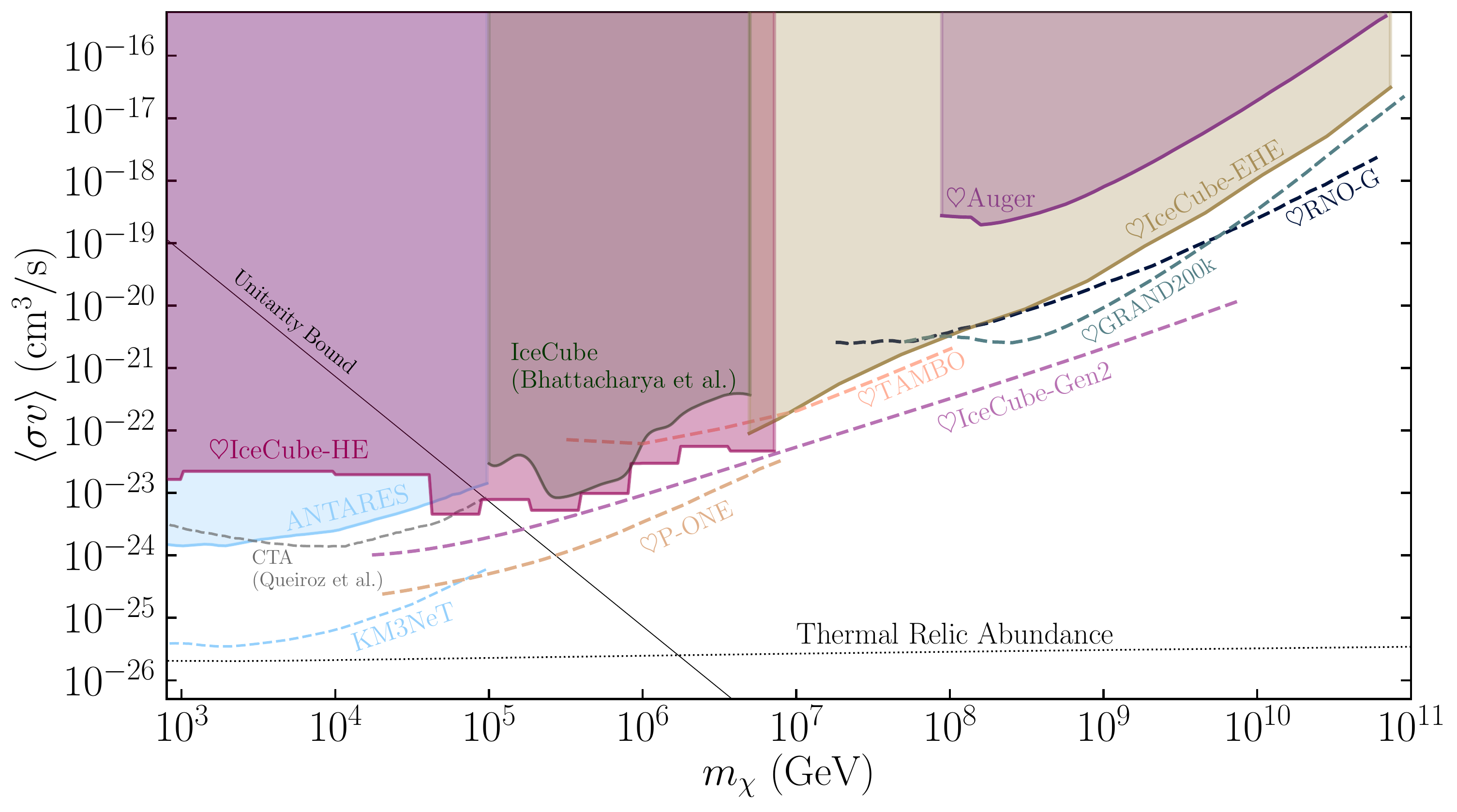}}
\caption{
\textbf{\textit{The landscape of supra-TeV dark matter annihilation into neutrinos.}}
Same as Fig.~\ref{fig:Indirect}, but for the high-mass region.
All the experimental constraints in this plot are calculated by converting either the detected flux or the reported upper limit into a conservative upper bound on the DM annihilation cross section.
}
\label{fig:Indirect_non_unitarity}
\end{figure*}

Our main results are shown in Figs.~\ref{fig:Indirect}-\ref{fig:Indirect_dwave}.
Fig.~\ref{fig:Indirect} shows the results derived according to the procedures described in Secs.~\ref{sec:galactic} and~\ref{sec:extragalactic}, in addition to previous results available in the literature.
Fig.~\ref{fig:Indirect_lowe} shows a more detailed view of the low-mass (sub-GeV) range; Fig.~\ref{fig:Indirect_non_unitarity} shows results for the high-mass (10$^3$-10$^{11}$ GeV) region.
Finally, Figs.~\ref{fig:Indirect_pwave}-\ref{fig:Indirect_dwave} provide the constraints and projections in the case of velocity-dependent p-wave and d-wave annihilation, respectively.
We label the results derived specifically for this work with a heart ($\heartsuit$). 

\renewcommand{\arraystretch}{1.5}

In the rest of this section, we describe the data that we used to produce or recast limits on DM annihilation into neutrinos according to the procedures outlined in Sec.~\ref{sec:theory}.
We split the data into three lists: 1) data used to construct constraints in Fig.~\ref{fig:Indirect}; 2) previous limits that we have recast; and 3) data used to place limits in the high mass ($m_\chi > 10^3$ GeV) region.

When reporting literature results, where possible, we have rescaled them to use the same halo parameters, \textit{i.e.} consistent $J$-factors, as computed in Sec.~\ref{sec:galactic}.
In this way, we ensure that the constraints we present can be properly compared one with another.
The rescaling could not be done in the case of ANTARES~\cite{Adrian-Martinez:2015wey}, SK~\cite{Frankiewicz:2017trk}, and IceCube~\cite{Aartsen:2016pfc}, since these were event-by-event analyses for which data is not publicly available.
This is unfortunate since the halo parameters used in these studies are no longer preferred  (see discussion in Sec.~\ref{sec:haloparams}).
Shaded regions correspond to experimental limits, whereas dashed lines are projections based on future experimental sensitivity.
Finally, we include two lines for reference.
First, the dotted black line corresponds to the cross section required to produce the observed relic abundance from thermal freeze-out computed as in \citet{Steigman:2012nb}, and second, the solid black line labeled ``unitarity bound'' corresponds to the perturbative unitarity limit on non-composite WIMP dark matter~\cite{Griest:1989wd}; see~\cite{Smirnov:2019ngs} for a recent discussion.

The limits shown in Fig.~\ref{fig:Indirect}, employing the approach of Secs.~\ref{sec:galactic} and ~\ref{sec:extragalactic}, use the following data, which we also summarize in Tbl.~\ref{tab:table3}. 
\begin{enumerate}
    \item \textbf{\textit{Borexino:}} Borexino is a large-volume unsegmented liquid scintillator detector located underground at the \textit{Laboratori Nazionali del Gran Sasso} in Italy~\cite{Alimonti:2008gc}. 
    The collaboration has released two event selections: one which has a livetime of 736 days selecting electron-antineutrino candidate events over the entire fiducial volume and another one with 482 days of livetime designed to search for geo-neutrinos~\cite{Bellini:2010hy}.
    These event selections are combined into a single set designed to obtain a pure sample of electron-antineutrinos by means of searching for signatures of inverse beta decay.
    Using this selection, they derive upper limits on the all-sky monochromatic electron-antineutrino flux ranging from $\sim10^5$ to $\sim10^2$ $\bar\nu_e{\rm cm}^{-2}{\rm s}^{-1}$, for energies ranging from $\sim2$ to $17$ MeV, respectively.
    We use the flux upper limits produced by ~\citet{Bellini:2010gn} and recently updated by \citet{Agostini:2019yuq} and compare it with one-sixth of the all-flavor expected flux from dark matter to set our constraints.
    \item \textbf{\textit{SNO+ (not shown):}} SNO+, located at the SNOLAB underground facility in Sudbury, Canada, consists of a 12m diameter acrylic vessel that will ultimately be filled with 780 tonnes of liquid scintillator and 800 kg of $^{130}$Te, with the goal of searching for neutrinoless double-beta decay~\cite{Andringa:2015tza}. Recent measurements in the water phase of SNO+ searching for invisible proton decay channels have been performed~\cite{Anderson:2018byx}.
    The event selection of this analysis looks for an atomic de-excitation into two gammas prompted by proton decay for a period of 114.7 days.
    For energies below $\sim$6 MeV the observed rate is well described by internal backgrounds produced by $^{214}$Bi and $^{208}$Ti decay chains; at higher energies they are dominated by electron-antineutrinos from nearby nuclear reactors interacting with atomic electrons.
    Neutrinos produced by dark matter can induce a similar signal when they have neutral current interactions with the medium.
    We computed the distribution of electron recoils in neutrino-electron charged-current interactions~\cite{rqft74, Formaggio:2013kya} and compared the expected rate to the observed sample rate given in~\cite{Anderson:2018byx}.
    The resulting limits from 5 to 30~MeV, assuming 100\% electron detection efficiency, lie above $\sv \gtrsim 10^{-20}$ cm$^{3}$s$^{-1}$.
    We do not include this line in our figures as inclusion of realistic efficiencies, which are not publicly available, will push these limits up. Depending on the tellurium-loading schedule, an extended scintillator-only run could substantially improve these limits.
    \item \textbf{\textit{KamLAND:}} KamLAND is an unsegmented liquid scintillator detector located in the Kamioka observatory near Toyama, Japan.
    The approximately one kiloton of mineral oil fiducial volume is contained in a 13 meter balloon.
    Beyond its well-known work on reactor neutrinos, KamLAND has measured the $^{8}$B solar spectrum~\cite{Abe:2011em}, searched for geoneutrinos~\cite{Gando:2013nba}, and placed limits on the flux of extraterrestrial neutrinos above $\sim 8.3~{\rm MeV}$~\cite{Collaboration:2011jza} which constrains the supernovae relic neutrino flux. 
    In the latter work, an upper limit on the extraterrestrial flux of $\bar\nu_e$ is derived, which is at the $\mathcal{O}(10)~\bar\nu_e~{\rm cm}^{-2}{\rm s}^{-1}{\rm MeV}^{-1}$ level and is given from 8.3 MeV to 18.3 MeV.
    Using this result, we derive a constraint on the dark matter annihilation into neutrinos, shown in salmon in Fig.~\ref{fig:Indirect}.
    Note that in~\cite{Collaboration:2011jza}, the KamLAND collaboration also derives a similar constraint, but with outdated $J$-factors; their result and ours are comparable.
    These are the leading constraints in the $\sim$10 MeV mass range, but we expect that they will be improved by the next-generation liquid scintillator detector in China, JUNO~\cite{An:2015jdp}.
    \item \textbf{\textit{SK:}} Super-Kamiokande (SK) is a 50kt ultrapure water Cherenkov detector located in Kamioka, Japan~\cite{Fukuda:2002uc}. 
    SK can use the morphology of the Cherenkov ring produced by charged particles to perform particle identification, energy measurement, and obtain directional information of the events.
    The unfolded electron- and muon-neutrino fluxes in the sub-GeV to several TeV energy range has been published by SK~\cite{Richard:2015aua}.
    This unfolding uses data from the four stages, SK-I, SK-II, SK-III, and SK-IV, resulting in a total livetime of 4799 days for the fully contained and partially contained event selection and 5103 for the upward-going muon sample. 
    The unfolded fluxes are expected to be dominated by the atmospheric neutrino flux; in fact they are in agreement with model predictions, {\it e.g.} the HKKM model~\cite{Honda:2006qj}, within systematic uncertainties.
    The dominant source of uncertainties on the unfolded fluxes is the neutrino interaction cross section, which introduces an uncertainty of approximately 20\% in the unfolded flux.
    In the case of electron-neutrinos, the second largest uncertainty is due to the small statistics at high energies; which can be up to 10\% in the highest energy bins. 
    For all flavors, all other sources of uncertainty are less than 5\% across all energy bins. 
    We compare the unfolded flux with the expected flux from dark matter to produce limits on Galactic and extragalactic dark matter annihilation. These results are shown in purple in Figs.~\ref{fig:Indirect},~\ref{fig:Indirect_pwave}, and~\ref{fig:hmfparam}, and labeled as \textit{\textbf{$\heartsuit$SK-Atm.}}
    In order to obtain these limits we used a background-agnostic approach as described in Sec.~\ref{sec:statistics}, and a binned truncated Gaussian likelihood in energy with two degrees of freedom.
    This result is complementary with SK Galactic dark matter annihilation analysis~\cite{Frankiewicz:2017trk,FRANKIEWICZ:2018lsh, Abe:2020sbr}, shown in teal in Fig.~\ref{fig:Indirect} and simply labeled \textit{\textbf{SK}}.
    As expected, our limits using the background agnostic method are weaker than ones produced by the collaboration, but our analysis extends to lower energy and covers the energy range from 0.1 to 100~GeV in dark matter mass.
    Additionally, we perform an analysis using 2853 days of low energy data from SK I/II/III, as well as 2778 days of data from SK phase IV, which led to an upper limit on the relic supernova electron antineutrino ($\bar \nu_e$) flux~\cite{WanLinyan:2018}; labeled \textit{$\heartsuit$\textbf{SK-$\mathbf{\bar \nu_e}$}}.
    The resulting limits on $\sv$ turn out to be the strongest over the entire mass range that we consider, flirting with the relic abundance line for masses between 27 and 30~MeV.
    \item \textbf{\textit{IceCube}}: The IceCube Neutrino Observatory is a gigaton ice Cherenkov neutrino detector located at the geographic South Pole~\cite{Aartsen:2016nxy}.
    IceCube has measured the atmospheric neutrino spectrum in the 100~GeV to 100~TeV energy range.
    By separating the events into their observed morphologies (``cascades'' and ``tracks''), the collaboration recently published the unfolded electron- and muon-neutrino flux in this energy range~\cite{Aartsen:2015xup,Aartsen:2016xlq}.
    At energies greater than 60~TeV, using events whose interaction vertex starts in the inner part of the detector~\cite{Aartsen:2013jdh,Schneider:2019ayi}, they have also reported the result of a piece-wise power-law fit to the astrophysical neutrino component using more than six years of data~\cite{Aartsen:2017nbu}.
    We use these to produce background-agnostic limits on the velocity averaged dark matter annihilation cross section by comparing the produced neutrino flux with the reported unfolding or spectral fits.
    The obtained limits are shown for dark matter masses from 200~GeV to 10~PeV, labeled \textit{\textbf{$\heartsuit$IceCube-HE}} and colored in dark magenta.
    Limits use the same likelihood construction as in the case of the SK limits described above. Note that the muon neutrino atmospheric unfolding reported by IceCube uses northern tracks, which are unfortunately in the wrong hemisphere for the Galactic center. Therefore, for that sample, we only constrain extragalactic emission.
    Dedicated neutrino line searches have not been yet performed by the IceCube collaboration, although sensitivities have been estimated in~\cite{ElAisati:2017ppn,ElAisati:2018vkn} to be stronger than current IceCube constraints in that region.
    We describe the region labeled \textit{\textbf{IceCube-EHE}} below, in the description of the high-mass region.
\end{enumerate}

\noindent Additionally, we use the following previously-published limits on dark matter annihilation obtained by constraining the Galactic flux, rescaled to account for the galactic halo parameters used here unless indicated otherwise:
\begin{enumerate}
    \item \textbf{\textit{Super-Kamiokande diffuse supernovae flux search}}: The gray region labeled \textit{\textbf{SK Olivares et al.}} is an independent analysis of SK all-sky low-energy data which uses SK phases I through III to derive an upper bound on the supernova relic neutrinos~\cite{Hosaka:2005um,Cravens:2008aa,Abe:2010hy}. 
    This analysis covers neutrino energies from 10 MeV to 200 MeV; see~\cite{Li:2014sea} for a recent discussion of backgrounds in the low-energy range.
    The upper limit on supernova relic neutrinos was then converted into dark matter annihilation constraints, and was originally presented in~\cite{Campo:2017nwh,Campo:2018dfh,Olivares-Del-Campo:2019qwe}.
    Recently, SK phase-IV data has placed new constraints on the $\bar\nu_e$ flux in the 10 to 30~MeV energy range~\cite{WanLinyan:2018}.
    These observations improve over KamLAND constraints~\cite{Collaboration:2011jza} by a factor between 3 and 10 in their overlapping energy range.
    Thus these observations dominate the constraints for dark matter masses below $\sim 20~{\rm MeV}$.
    Where they overlap, the \textit{Olivares et al.} limits are not quite as strong as the \textit{SK-$\bar \nu_e$} limits that we have presented, because their background modelling could not use angular information which is not publicly available.
    \item \textbf{\textit{Super-Kamiokande Galactic dark matter search}}: The teal region, labeled \textit{\textbf{SK}}, is from~\cite{Frankiewicz:2015zma}.
    This analysis uses muon-neutrino data in the energy range between 1 GeV and 10 TeV collected by SK over 5325.8 days.
    Since this analysis relies on angular information that is not public, it has not been rescaled to account for our choice of galactic halo parameters.
    \item \textbf{\textit{IceCube/DeepCore Galactic dark matter search}}: The IceCube limits are from~\cite{Aartsen:2016pfc} and use 329 days of IceCube data.
    These  place constraints for masses in between 25~GeV and 10~TeV.
    At the lowest masses, these limits include data from DeepCore, an array of more closely spaced inner strings in  IceCube.
    In addition, we include a limit derived from 3 years of data using primarily tracks to constrain Galactic center emission~\cite{Aartsen:2017ulx}.
    For display purposes, we join these two lines, choosing the best limit at each point, and show it in navy blue, simply labeled as \textit{\textbf{IceCube}}.
    \item\textbf{\textit{IceCube-Bhattacharya et al.}} is taken from ~\cite{Bhattacharya:2019ucd}'s channel-by-channel unbinned likelihood analysis of the High-Energy Starting Event (HESE) data, including energy, angular, and topology information.
    They include both Galactic and extragalactic constraints.
    Constraints that we derive (IceCube-HE) using only spectral information follow these limits quite closely at higher energies since the small sample size prevent angular information from contributing significantly.
    \item \textbf{\textit{ANTARES dedicated Galactic dark matter search}}: The light blue region, labeled ANTARES, is from a Galactic center analysis of nine years of ANTARES muon neutrino and antineutrino data~\cite{Albert:2016emp,Adrian-Martinez:2015wey}. This covers the dark matter mass range from $53$ GeV to $100$ TeV.
    \item \textbf{\textit{Baikal dedicated Galactic dark matter search} (not shown)}: The Baikal underwater neutrino telescope~\cite{Belolaptikov:1997ry,Aynutdinov:2006ca}, NT-200, is a water Cherenkov detector deployed in Lake Baikal, Russia.
    It has an instrumented volume of approximately 100~kt and is comprised of 192 optical modules arranged on eight strings, with a typical distance between strings of 21~m.
    The collaboration performed an analysis looking for dark matter annihilation in the Galactic center into neutrinos using data recorded between April of 1998 to February of 2003~\cite{Avrorin:2015bct}.
    This analysis claimed to place limits on the cross section at the $10^{-22}~{\rm cm}^3 {\rm s}^{-1}$ level for a 1~TeV dark matter mass.
    We do not add this result to our constraint summary because there are stronger results in this mass range, but we do show the projections of the next generation detector at Lake Baikal, GVD.
    \item \textbf{\textit{Combined IceCube and ANTARES dedicated Galactic dark matter search} (not shown)}: Recently~\citet{Aartsen:2020tdl} have performed a combined analysis of the IceCube and ANTARES data sets which corresponds to approximately 1000 days of the former and 2000 of the latter. 
    The combined result only marginally improves the previously published results, which we include in this review.
    The most notable point of this work is the consideration of underfluctuations when placing constraints on the data.
    In previous work by ANTARES, when the obtained data limit exceeds the mean sensitivity the reported result was the sensitivity of the analysis, while in the previous IceCube work underfluctuations are taken into account in the statistical limit and reported.
    Given an underfluctuation of data observed in the ANTARES data set, the combined result is  approximately a factor of two stronger in the ANTARES dominated region.
    We do not show the results of this analysis in our plot summary for two reasons: the analysis only reports the experiment-overlapping dark matter parameter range from 50~GeV to 1~TeV and does not report the $\nu \bar \nu$ channel that we study in this work.
\end{enumerate}
Finally, Fig.~\ref{fig:Indirect}, includes next-generation sensitivities that can be reached by future experiments. These are shown as dashed lines:
\begin{enumerate}
\item \textbf{\textit{DUNE}}: The Deep Underground Neutrino Experiment (DUNE) far detector will be a 46.4 kiloton liquid argon Time Projection Chamber (TPC)~\cite{Acciarri:2015uup,Abi:2020wmh} constructed at the Sanford Underground Research Facility (SURF) in South Dakota, USA.
Its main advantage in detecting neutrinos from DM annihilation is its improved particle identification, using morphological reconstruction, with respect to Cherenkov detectors like Super-Kamiokande, ANTARES, or IceCube, which \textit{e.g.} can be exploited to make improved measurements of solar neutrinos~\cite{Capozzi:2018dat}.
Thus, a dedicated DUNE analysis utilizing the expected improved directional capability can prove effective in a search for Galactic dark matter annihilation to neutrinos.
We derive projected sensitivities for dark matter masses in the range from 100~MeV to 30~GeV and show them in Fig.~\ref{fig:Indirect} as dashed orange lines.
The dominant background in this energy range is from atmospheric neutrinos.
We use the predictions provided by \citet{PhysRevD.92.023004} at the Homestake gold mine at SURF, taking into account oscillations through the Earth using the nuSQuIDS package~\cite{Delgado:2014kpa,Arguelles:2020hss,nusquids}.
In our analysis, we consider $e$- and $\tau$-flavored charged-current interactions and compare the expected energy distribution; \textit{i.e.} we do not take into account event-by-event directional information.
We use a fractional charged lepton energy resolution of $2\% + 15\%/\sqrt{E/{\rm GeV}}$~\cite{Acciarri:2015uup} and assume the idealized condition of $100\%$ efficiency.
In our analysis, charged-current electron-neutrino interactions are assumed to deposit all their energy in the detector, while tau-neutrino charged-current interactions will deposit less visible energy due to the invisible neutrinos produced in the prompt $\tau$ decay.
Since we expect that DUNE morphological identification will be able to single out muon-neutrino charged-current processes, we choose to remove them from the analysis as they are the primary contributor to the atmospheric neutrino background.
Limits are derived using a binned Poisson likelihood and a background-informed method as described in Sec.~\ref{sec:statistics}.
We note that, due to liquid argon TPC's morphological reconstruction capabilities, a proper Galactic center analysis including directionality would benefit from the inclusion of muon-neutrino charged-current interactions, and thus our projections are conservative.
\item \textbf{\textit{Hyper-Kamiokande}}: Building on SK's technology, a new water Cherenkov detector with a fiducial mass of 187~kton called Hyper-Kamiokande (HK) will be built in Kamioka, Japan~\cite{Abe:2018uyc}.
Due to its larger size, this detector will be able to place stronger limits on the DM annihilation cross section to neutrinos than its predecessor~\cite{Campo:2018dfh}.
In fact, Hyper-Kamiokande is estimated to reach $\sim 10^{-25}~{\rm cm}^3 {\rm s}^{-1}$ for $1~{\rm GeV}$ dark matter and $\sim 10^{-22}~{\rm cm}^3 {\rm s}^{-1}$ at $10^4~{\rm GeV}$ with ten years of data taking~\cite{Migenda:2017tej}.
Furthermore, the possibility of doping both the SK and the HK detectors with gadolinium (Gd) will reduce the dominant background for low-energy analyses by a factor of five and, consequently, improve the constraints on DM annihilation~\cite{Horiuchi:2008jz,Laha:2013hva,Bell:2020rkw}.
\citet{Bell:2020rkw} performed a detailed directional analysis of DM annihilation in the MW, including Monte Carlo simulation of the atmospheric and diffuse supernova neutrino background as well as the detector geometry.
Figure~\ref{fig:Indirect} shows their equivalent results for five years of run time, which range from $\sv \lesssim 10^{-25}$ cm$^{2}$ at $m_\chi = 16$~MeV, to $\sv \lesssim 4.3 \times 10^{-24}$ cm$^{2}$ at 50~GeV. 
For the $p$- and $d$-wave constraints in Sec.~\ref{sec:pwave} we derive our own projected sensitivities for five years of data taking for DM masses in the 100~MeV to 30~GeV range, as the directional dependence does not allow the Bell~\textit{et al.} curve to be rescaled. 

Similar to our DUNE analysis, we assume that the dominant background in this energy range is due to atmospheric neutrinos, where we use the predictions provided by \citet{PhysRevD.92.023004} at the Kamioka mines, and allow these neutrinos to oscillate through the Earth using the nuSQuIDS package~\cite{Delgado:2014kpa,nusquids}.
We only consider $e$- and $\tau$-flavored charged-current interactions, without taking into account directionality.
We make the same assumptions as our DUNE analysis regarding energy deposition, while using an energy resolution of $1.5\% + 2\%/\sqrt{E / {\rm GeV}}$~\cite{Jiang:2019xwn}. We use total energy rather than lepton (visible) energy, which leads to a sensitivity overestimate of $\sim 40\%$ but simplifies the analysis. In principle, it is be possible to record lepton and proton energy above the proton Cherenkov threshold, \citep[see e.g.][]{Fechner:2009aa}.
We follow the same statistical procedure as in DUNE and, like DUNE, the sensitivity strength derives primarily from the expected electron- and tau-neutrinos signal.
Taking advantage of this channel explains why our estimates are better than ones presented by \citet{Migenda:2017tej}; see \citet{Beacom:2004jb} for a discussion on ``shower power.''
We have checked that the corresponding $s$-wave results agree well with Bell~\textit{et al.} below $\sim 1$ GeV within their quoted uncertainties. However, due to the incorporation of angular observables, enabled by their dedicated simulation, their limits are better by a factor of $\sim 2$ above $\sim 1$ GeV.
These projected sensitivities, especially at low energies, are subject to a $\sim 30\%$ uncertainty due to a combination of atmospheric background uncertainties and neutrino cross sections.
\item \textbf{\textit{JUNO}}: The Jiangmen Underground Neutrino Observatory~\cite{An:2015jdp} is a 20~kt unsegmented liquid scintillator detector under deployment in the Guangdong province of China.
The detector has a muon tracker on top of it and is also surrounded by water. 
Both of these systems can be used to veto cosmic-ray muons by either tagging them in the muon tracker or by detecting their Cherenkov light in water.
Due to its large volume and good energy resolution (estimated to be $3\%/\sqrt{E/{\rm MeV}}$) we expect that this experiment will have good sensitivity for neutrino line searches.
We estimate the sensitivity of JUNO to dark matter annihilation to neutrinos in the electron antineutrino channel following the proposal given in \citet{PalomaresRuiz:2007eu}.
We use background estimates derived for diffuse supernova background searches, as presented in \citet{An:2015jdp}.
Below 11 MeV, reactor antineutrinos dominate the background.
Between 11 and 40 MeV, the backgrounds are primarily neutral current interactions from atmospheric neutrinos, with sub-dominant charge current contributions.
According to our projection, JUNO is expected to constrain the velocity-averaged annihilation cross section better than $10^{-25}~{\rm cm}^3 {\rm s}^{-1}$ in the 10 to 40~MeV mass range.
The estimate is shown in dark red in Fig.~\ref{fig:Indirect}.

\item \textbf{\textit{INO} (not shown)}: The 50~kt magnetized Iron Calorimeter (ICAL)~\cite{Kumar:2017sdq,INOproc} at the India-based Neutrino Observatory is a planned segmented mille-feuille of iron plates interleaved with resistive plate chambers (RPCs).
The three modules will contain 151 iron leaves each, and a total of over 30,000 RPC units.
A 1.5~T magnetic field will allow discrimination between muon neutrinos and antineutrinos.
Following the successful completion of the mini-ICAL prototype, the INO underground laboratory and ICAL experiment are scheduled for construction at \textit{Pottipuram}, in the Bodi West hills of Theni District of Tamil Nadu, India.
~\citet{Khatun:2017adx} performed a forecast of the ICAL sensitivity to DM annihilation to neutrinos.
The ability to discriminate $\nu$ from $\bar \nu$ events provide a factor of 2-3 boost in sensitivity, which, when rescaled to 5 years,  ranges from $\sv \gtrsim 2 \times 10^{-24}$ cm$^{3}$ s$^{-1}$ at $m_\chi = 2$ GeV, to $10^{-23}$ cm$^{3}$ s$^{-1}$ at 90 GeV.

    \item \textbf{\textit{IceCube Upgrade}}: The IceCube Upgrade is an extension of the current IceCube/DeepCore array with seven closely-packed strings.
    These new strings will be separated by approximately 20 meters and each contain 100 photomultiplier tubes spaced vertically by 3~meters~\cite{Ishihara:2019aao}.
    Additionally, a number of calibration devices and sensors will be deployed to improve the modelling of the ice~\cite{Nagai:2019uaz,Ishihara:2019uei}.
In ~\cite{Baur:2019jwm} a preliminary estimation of the IceCube Upgrade sensitivity was performed. It is expected to be better than $10^{-24}~{\rm cm}^3 {\rm s}^{-1}$ for a 10~GeV dark matter mass.
\item \textbf{\textit{IceCube Gen-2}}: The next-generation ice Cherenkov neutrino observatory in Antarctica is a substantial expansion to the current IceCube observatory, aiming at enhancing the detector volume by a factor of ten~\cite{Aartsen:2014njl}.
This increased effective area is expected to provide a better sensitivity to resolve sources of high-energy cosmic neutrinos and identify components of cosmic neutrino flux.
Dark matter annihilation limits from IceCube presented here should therefore scale by at least the increased sample size due to the larger effective area.
We have recast the estimates of diffuse flux sensitivity given in~\cite{Aartsen:2019swn} to estimate the sensitivity to dark matter annihilation.

\item \textbf{\textit{Baikal-GVD}}: The Baikal Gigaton Volume Detector (GVD) is a planned expansion to the existing NT-200 detector, and is currently being deployed in Lake Baikal, Russia.
The detector has recently reached an effective volume of $\sim0.35$~km$^3$ and has already seen first $\nu$-light~\cite{Avrorin:2019vfc}.
The full array will contain 10,386 optical modules divided among 27 clusters of strings, and is expected to have a final instrumented volume of around $1.5$~km$^3$.
The sensitivity of GVD to Galactic dark matter annihilation has been estimated in~\cite{Avrorin:2014vca} and is shown as a dashed brown line labeled \textbf{GVD}.
\item \textbf{\textit{KM3Net}}: The km$^3$-scale water Cherenkov detector currently under construction in the Mediterranean sea is designed to provide high-purity increased effective areas in the Southern Hemisphere.
The larger effective area and improved angular resolution, compared to ANTARES, are expected to provide better constraints on Galactic dark matter.
Two separate sites are under construction for low- and high-energy regimes~\cite{Adrian-Martinez:2016fdl}.
The high-energy site, called KM3NeT/ARCA, will consist of two detector array blocks located approximately 100~km offshore from \textit{Porto Palo di Capo Passero}, Sicily, Italy~\cite{Aiello:2018usb}.
Each block is expected to have 115 strings with an average spacing of 90~m.
The low-energy site, called KM3NeT/ORCA, consists of one array block and is under deployment approximately 40~km south of Toulon, France; close to the ANTARES site.
The array is made out of 115 strings with an average horizontal spacing of 20~m.
Each string contains 18 optical modules; in KM3NeT/ARCA they are spaced vertically by 36~m, while in KM3NeT/ORCA they are spaced 9~m.
The horizontal spacing and number of strings are proportional to the effective volume of the experiment, while the vertical spacing is related to the energy threshold~\cite{Halzen:2005qu}.
KM3NeT/ARCA's science program is mainly oriented towards higher-energy (astrophysical) neutrino searches, while KM3NeT/ORCA will measure neutrino oscillations using atmospheric neutrinos.
Assuming an $E^{-2}$ democratic-flavor astrophysical neutrino flux with a normalization of $\sim 1.8\times 10^{-8} {\rm GeV}^{-1}{\rm s}^{-1}{\rm cm}^{-1}{\rm sr}^{-1}$ and an exponential cut-off at 3~PeV they expect to see 11~$\nu_\mu$'s, 41~$\nu_e$'s, and 26~$\nu_\tau$'s in five years of KM3NeT/ARCA operation~\cite{Adrian-Martinez:2016fdl}.
In Fig.~\ref{fig:Indirect} we show the KM3NeT/ARCA expected sensitivity to dark matter annihilation to neutrinos in five years of data taking~\cite{gozzini2019search}.
Their sensitivity is within a factor of a few from the expected relic abundance cross section for dark matter masses around a TeV.
\item \textbf{\textit{P-ONE}}: The Pacific-Ocean Neutrino Experiment (P-ONE) is a newly proposed multi-cubic kilometer neutrino detector utilizing sea water as Cherenkov medium~\cite{Agostini:2020aar}.
P-ONE would be deployed in the Cascadia Basin, off the coast of Vancouver island in the Pacific Ocean, taking full advantage of the Ocean Network Canada infrastructure and expertise already in place.
The main goal of the experiment is to explore the origin of the extraterrestrial neutrino flux.
A pair of test strings, named STRAW~\cite{Bedard:2018zml}, has already been successfully deployed and has collected water absorption data.
The first phase of the detector, known as the Pacific Ocean Neutrino \textit{Explorer}, involving ten strings is planned to be deployed in 2023.
Each string is planned to be equipped with twenty photomultiplier tubes.
The full detector is expected to be complete by 2030 with 70 strings.
Projected limits include backgrounds from atmospheric and diffuse astrophysical neutrinos, and use the exposures shown in~\citet{Agostini:2020aar}.
\item \textbf{\textit{TAMBO}}: The Tau Air-Shower Mountain-Based Observatory is a proposed array of small water-Cherenkov tanks to be deployed on either the Colca Valley or Cotahuasi Canyon in Peru~\cite{Wissel:2019alx,Romero-Wolf:2020pzh}.
These are two of the world's four deepest valleys and their unique geometry allows for efficient detection of Earth-skimming PeV $\nu_\tau$.
Most of the Colca Valley runs along a North-South corridor, though a smaller section of it has an East-West corridor.
If deployed in the East-West corridor of the Colca valley, the declination band covered is $-15.5 \pm 10$ degrees, while in the North-South corridor it would be $-15.5\pm 50$ degrees.
These two provide two extreme configurations in terms of its GC exposure, while a deployment in the Cotahuasi canyon, which has an approximately diagonal corridor, would provide an intermediate exposure. 
TAMBO's effective area is expected to be 10 times larger than IceCube $\nu_\tau$~\cite{Aartsen:2013jdh} at a PeV and 30 times larger at 10 PeV.
The use of the Earth-skimming technique is complementary to very-high-energy Earth-traversing neutrino searches~\cite{Safa:2019ege} and the fact that it relies on the Cherenkov effect, rather than the higher energy threshold Askaryan effect, gives it unique potential to constrain dark matter in the tens of PeV mass range.
Depending on the final geometry of TAMBO its sensitivity to dark matter ranges from $10^{-22}~{\rm cm}^3~{\rm s}^{-1}$ to $4 \times 10^{-21}~{\rm cm}^3~{\rm s}^{-1}$ for a 1 PeV dark matter mass.
Sensitivities shown here are recast from the diffuse flux sensitivity presented by~\cite{Wissel:2019alx}.
A similar detector has been proposed to be deployed in Hawaii~\cite{2014arXiv1409.0477H, Sasaki:2017uta, 2019ICRC...36.1003S}.

\item \textbf{\textit{CTA}}: The Cherenkov Telescope Array is a planned network of 99 air Cherenkov telescopes in the southern hemisphere and 19 in the northern hemisphere that will collectively provide full-sky coverage of the gamma ray sky over an energy range from 20 GeV to 300 TeV~\cite{Acharya:2017ttl}.
Several CTA prototypes have been built and some have already seen first light. The telescopes are projected to have an angular resolution down to 0.1 degrees and a duty cycle of $\sim15\%$.
For high-mass dark matter annihilation into neutrinos, electroweak final-state radiation can also lead to the production of gamma rays, despite a completely ``invisible'' $\nu \bar \nu$ final state, and can thus be constrained by gamma ray observations of the Galactic center with CTA; see Sec.~\ref{sec:theory} for more details.
The expected limits from CTA were computed in~\cite{Queiroz:2016zwd}, and shown as a dashed silver line assuming 100 hours of observation.
\end{enumerate}

We note that the 10 MeV -- 1 GeV range can in principle be covered by future tonne-scale dark matter direct detection experiments such as DARWIN and ARGO~\cite{McKeen:2018pbb}.
However, these are still in their planning phases, meaning that construction is still decades away, and very long ($\gtrsim$ 10 years) exposure times are required to be competitive with HyperK. For this reason we do not show them here. 

Fig.~\ref{fig:Indirect_non_unitarity} shows the extension of available constraints to larger masses, above the ``unitarity bound,'' accessible \textit{e.g.} for composite DM models~\cite{Frigerio:2012uc}.
These bounds are calculated by converting either the detected flux or reported upper limits, from observatories sensitive to these mass range, into a conservative upper bound on the DM annihilation to neutrinos.  The following experiments are sensitive to this regime:
\begin{enumerate}
\item \textbf{\textit{Auger}}: The Pierre Auger Observatory is a hybrid detector consisting of both an array of water Cherenkov surface detectors and atmospheric fluorescence detectors.
Located in Malarg\"ue, Argentina~\cite{ThePierreAuger:2015rma} and operational since 2004, the collaboration has made a multitude of measurements of the highest energy cosmic rays.
This includes measurements of the spectral distribution of cosmic rays beyond the GZK limit, anisotropy searches, as well as fits to their mass composition.
Beyond the extensive cosmic ray program, Auger is able to probe extremely-high-energy neutrinos by searching for showers developing deep in the atmosphere, since showers induced by cosmic rays are likely to develop much earlier.
Another possible detection channel is upgoing tau lepton showers, which are induced by Earth-skimming tau neutrino interactions near Earth's surface.
In 2017, the collaboration reported a limit on the diffuse flux of high energy neutrinos between $10^{8}-10^{11}$ GeV~\cite{Zas:2017xdj} which we use to set a background-agnostic bound on $\sv$ for such energies (purple line in Fig.~\ref{fig:Indirect_non_unitarity}).
\item \textbf{\textit{IceCube-EHE}}: Beyond the astrophysical neutrino flux, IceCube performs searches for GZK neutrinos using a dedicated sample of events that deposit extremely high energies (EHE) in the detector.
The most recent search used nine years of data and set limits on the GZK flux.
We use these limits~\cite{Aartsen:2018vtx} to derive an upper bound on the DM annihilation cross section to neutrinos between $10^{7}-10^{11}$ GeV, represented by a light brown line in Fig.~\ref{fig:Indirect_non_unitarity}.
\item \textbf{\textit{ANITA (not shown)}}: The ANtarctic Impulsive Transient Antenna is an array of radio antennas attached to a helium balloon that flies for $\sim30$ days at a time above Antarctica.
The goal of this experiment is to measure the GZK (cosmogenic) neutrino flux by detecting radio showers emitted by extremely-high-energy neutrinos after interacting in the Antarctic ice~\cite{Gorham:2008dv}.
The collaboration has successfully completed four such flights, setting the strongest limits on astrophysical neutrino fluxes above $10^{11}$~GeV; anomalies notwithstanding.
We derive limits on dark matter annihilation to neutrinos by rescaling the reported upper limits from the fourth flight of ANITA~\cite{Gorham:2019guw}.
They extend up to $m_\chi = 10^{12}$~GeV, but do not constrain $\sv$ to be any smaller than $10^{-14}$ cm$^3$s$^{-1}$, putting them outside of the range of Fig.~\ref{fig:Indirect_non_unitarity}..
\item \textbf{\textit{GRAND}}: The Giant Radio Array for Neutrino Detection is a proposed large-scale observatory consisting of 200,000 radio antennas covering 200,000 km$^2$ near a mountain range in China.
This experiment plans to use the surrounding mountains as a target for Earth-skimming tau neutrinos.
After the neutrinos interact in the mountain, a tau lepton should be observed exiting the mountain and subsequently decaying in the atmosphere.
The immense coverage will allow GRAND to probe GZK neutrino fluxes that are at least an order of magnitude below current limits~\cite{Alvarez-Muniz:2018bhp}.
We convert their 3-year sensitivity to the GZK neutrino flux between $10^{8}-10^{11}$ GeV into sensitivities on $\sv$ shown as a dashed navy blue line in Fig.~\ref{fig:Indirect_non_unitarity}.

\item \textbf{\textit{RNO-G}}: The Radio Neutrino Observatory in Greenland aims to measure the neutrino flux above $10^{16}$~eV~\cite{Aguilar:2019jay}.
The array of antennas to be deployed in the ice are designed to detect the Askaryan radio emission from extremely high-energy neutrinos traversing the Earth and atmosphere.
The design and deployment of RNO relies upon the experience and expertise obtained in successful deployment and operation of ARA and ARIANNA~\cite{2012APh....35..457A,Barwick:2014pca}.
The plan is to deploy 35 stations such that each station will consists of a surface array and a deep array.
The surface array is going to be used for cosmic-ray detection while the deep array, benefiting from a large effective volume, will detect neutrinos.
\item \textbf{\textit{BEACON} (not shown)}: Beamforming Elevated Array for Cosmic Neutrinos is another experiment proposed to search for the flux of very high energy neutrinos beyond 100 PeV.
An array of antennas installed at high elevations and presumes the use of a beamformer radio array. 
The project is currently in prototype stage, being tested at the White Mountain Research Station in California~\cite{Wissel:2020sec}.
The Cotahuasi Canyon, where TAMBO is deployed, has been considered as a potential site for BEACON.
Given that the site of BEACON is yet to be confirmed, we have not projected the sensitivity for it in this review.

\item \textbf{\textit{POEMMA (not shown)}}: The Probe Of Extreme Multi-Messenger Astrophysics is a proposed probe-class space mission to observe ultra-high-energy cosmic rays and neutrinos above 20 PeV.
Two satellites on near-equatorial orbits will observe fluorescence caused by showers in the Earth's atmosphere.
When in stereo observation mode, POEMMA will effectively monitor $10^{13}$ metric tons of atmosphere~\cite{Anchordoqui:2019omw,Olinto:2020oky}.
Preliminary diffuse neutrino flux sensitivity studies have projected as much as an order of magnitude improvement over existing limits at energies greater than $10^{10}$~GeV.
We do not include POEMMA here, as neutrino sky coverage maps were not available at the time of this analysis. 
\end{enumerate}

\begin{table*}[!ht]
\begin{center}
{
\begin{tabular}{ |c|m{18em} |c|c|m{10em}| } 
 \hline
 \textbf{Energy Range} & \textbf{Experimental Analysis} & \textbf{Directionality} & \textbf{Detected Flavor} \\ \hline 
 $2.5-15$ MeV & Borexino \cite{Bellini:2010gn}& $\boldsymbol{\times}$ &  $\bar{\nu}_e$ (IBD) \\ \hline
 $8.3-18.3$ MeV & KamLAND \cite{Collaboration:2011jza} &  \checkmark &  $\bar{\nu}_e$ (IBD) \\\hline
 {$10-40$ MeV} & JUNO \cite{An:2015jdp} &  \checkmark &  $\bar{\nu}_e$ (IBD) \\\hline
  \multirow{2}{*}{ $15-10^3$ MeV} 
  & SK \cite{Campo:2017nwh} & $\boldsymbol{\times}$  &  $\bar{\nu}_e$ (IBD)\\ \cline{2-4}
  & DARWIN \cite{McKeen:2018pbb} & $\boldsymbol{\times}$  &  All Flavors (Coherent)\\\hline
  $0.1-30$ GeV & DUNE \cite{Abi:2020evt} \newline HK \cite{Campo:2018dfh} & $\boldsymbol{\times}$ &  $\nu_e, \bar{\nu}_e, \nu_{\tau}, \bar{\nu}_{\tau}$ (CC)\\\hline
  $1-10^4$ GeV & SK \cite{Frankiewicz:2015zma, Abe:2020sbr} & \checkmark &  All Flavors \\\hline
    $20-10^4$ GeV & IceCube \cite{Aartsen:2016pfc} & \checkmark &  All Flavors\\\hline
     $50-10^5$ GeV & ANTARES \cite{Adrian-Martinez:2015wey} & \checkmark &  $\nu_\mu,\,\bar{\nu}_\mu$ (CC)\\\hline
         $0.2-100$ TeV & CTA \cite{Queiroz:2016zwd} &  \checkmark &  All Flavors (Bremsstrahlung)\\\hline
          $10 - 10^4$ GeV & IC-Upgrade~\cite{Baur:2019jwm} &  \checkmark &  All Flavors \\\hline
          $> 10$ PeV & IC~Gen-2~\cite{Aartsen:2014njl} &  \checkmark &  All Flavors \\\hline
          $10 - 10^4$ TeV & KM3Net~\cite{Adrian-Martinez:2016fdl} &  \checkmark &  All Flavors \\\hline
          $1 - 100$ PeV & TAMBO~\cite{Wissel:2019alx} &  \checkmark &  $\nu_\tau,\,\bar{\nu}_\tau $ (CC) \\\hline
          $> 100$ PeV & GRAND~\cite{Alvarez-Muniz:2018bhp} &  \checkmark &  $\nu_\tau,\,\bar{\nu}_\tau $ (CC) \\\hline
\end{tabular}
}
\caption{
\textbf{\textit{Summary of current and future experiments discussed in this work for different energy ranges.}}
The table also indicates whether the experimental analysis used directional information and which neutrino flavors it relied on.
}
\label{tab:table3}
\end{center}
\end{table*}

\subsection{Velocity-dependent annihilation}

\begin{figure*}[ht!]
\centerline{\includegraphics[width=\linewidth]{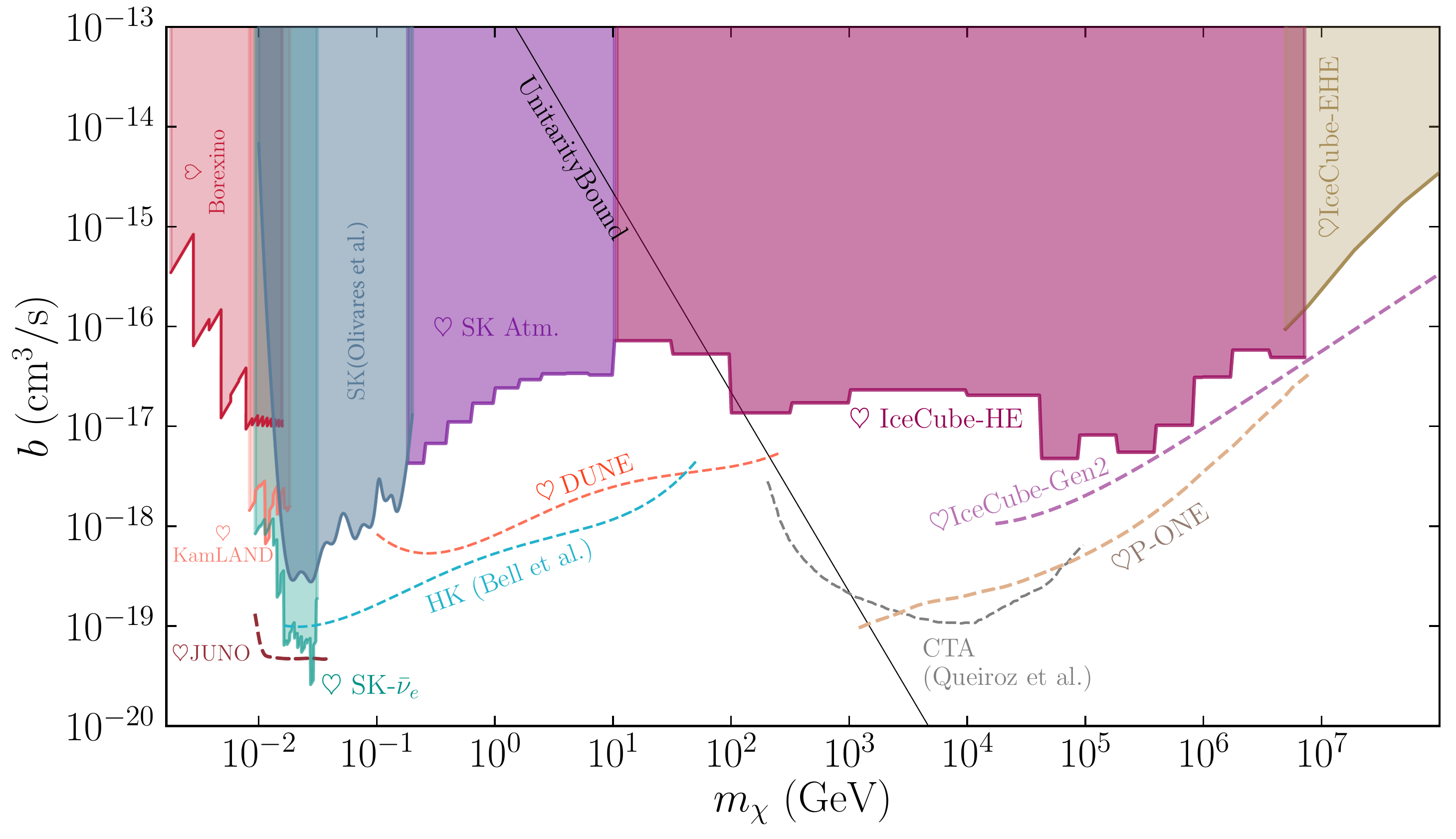}}
\caption[]{Limits on $p$-wave, $\sv = b(v/c)^2$, velocity-dependent annihilation cross-section of dark matter to two neutrinos. The cross section needed to explain the observed abundance for thermal DM is $\langle \sigma v_r \rangle = 6\times 10^{-26}\; \rm{cm}^3/\rm{s}$.}
\label{fig:Indirect_pwave}
\end{figure*}

\begin{figure*}[hbt!]
\centerline{\includegraphics[width=\linewidth]{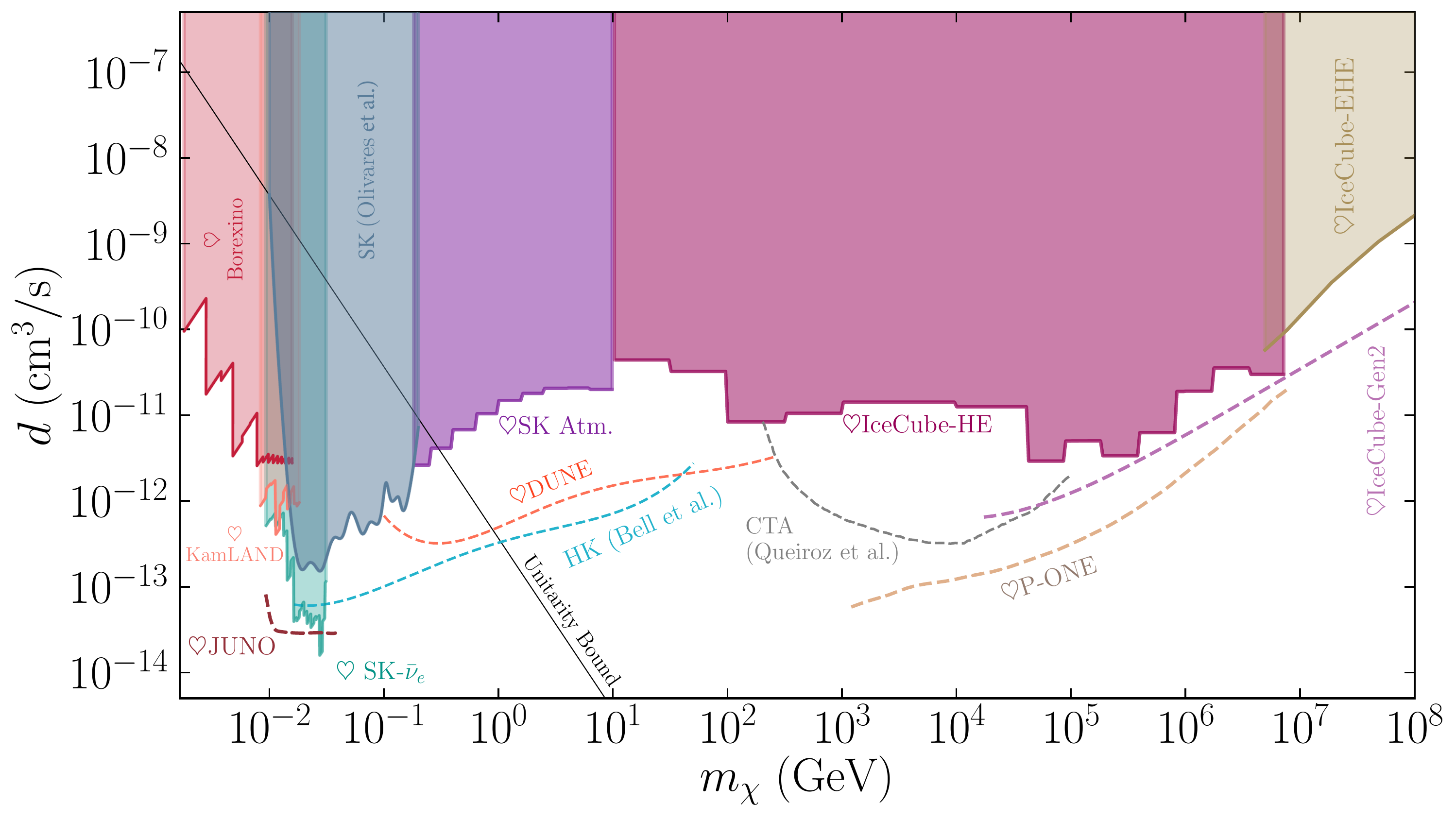}}
\caption[]{
Limits on the annihilation of neutrinos to dark matter through a $d$-wave process $\langle \sigma v \rangle = d (v/c)^4$.
}
\label{fig:Indirect_dwave}
\end{figure*}

Fig.~\ref{fig:Indirect_pwave} shows the corresponding limits for $p$-wave annihilation, and Fig.~\ref{fig:Indirect_dwave} provides limits on $d$-wave annihilation.
In these cases, we follow the procedures outlined in Sec.~\ref{sec:pwave}, to reweight the astrophysical portion of the flux prediction (Eqs.~\eqref{eq:galaxyAnnRate} and~\eqref{eq:overdensity}) to account for the dark matter velocity dispersion.
We do this for all-sky searches since analyses where the angular distribution of the neutrinos has been taken into account are not easily re-scaled when considering the velocity distribution of DM particles within the halo.
Similarly, all the constraints taken from the literature are re-scaled using our choice of halo parameters (see Tbl.~\ref{tab:Jtable} for halo parameters and $J$-factor for the different analyses in the literature). Unsurprisingly, the limits on $\sv$ are much weaker for $p-$ and $d-$wave processes due to the strong velocity suppression. In contrast to the $s-$wave case, where the smallest halos tend to dominate the expected signal, velocity-suppressed annihilation is strongest in the largest DM halos where dispersion velocities are higher. These limits are thus insensitive to the value of the minimum halo mass $M_{min}$. However, the constraints from annihilation in the Milky Way halo remain dominant over the extragalactic contribution. 

\subsection{Dark matter halo uncertainties\label{sec:haloparams}}

As previously mentioned, a major source of uncertainty comes from the spatial dark matter distribution, because of the $n_\chi^2$ dependence in the annihilation signal.
For Galactic constraints, this is mainly reflected by uncertainties in the Milky Way dark matter distribution.
For extragalactic constraints, we focus on the shape of the halo mass function and the minimum dark matter mass, which determines how far down extrapolations of the HMF must go to account for the total DM contribution.

\begin{figure*}[t!]
    \centering
    \includegraphics[width=\linewidth]{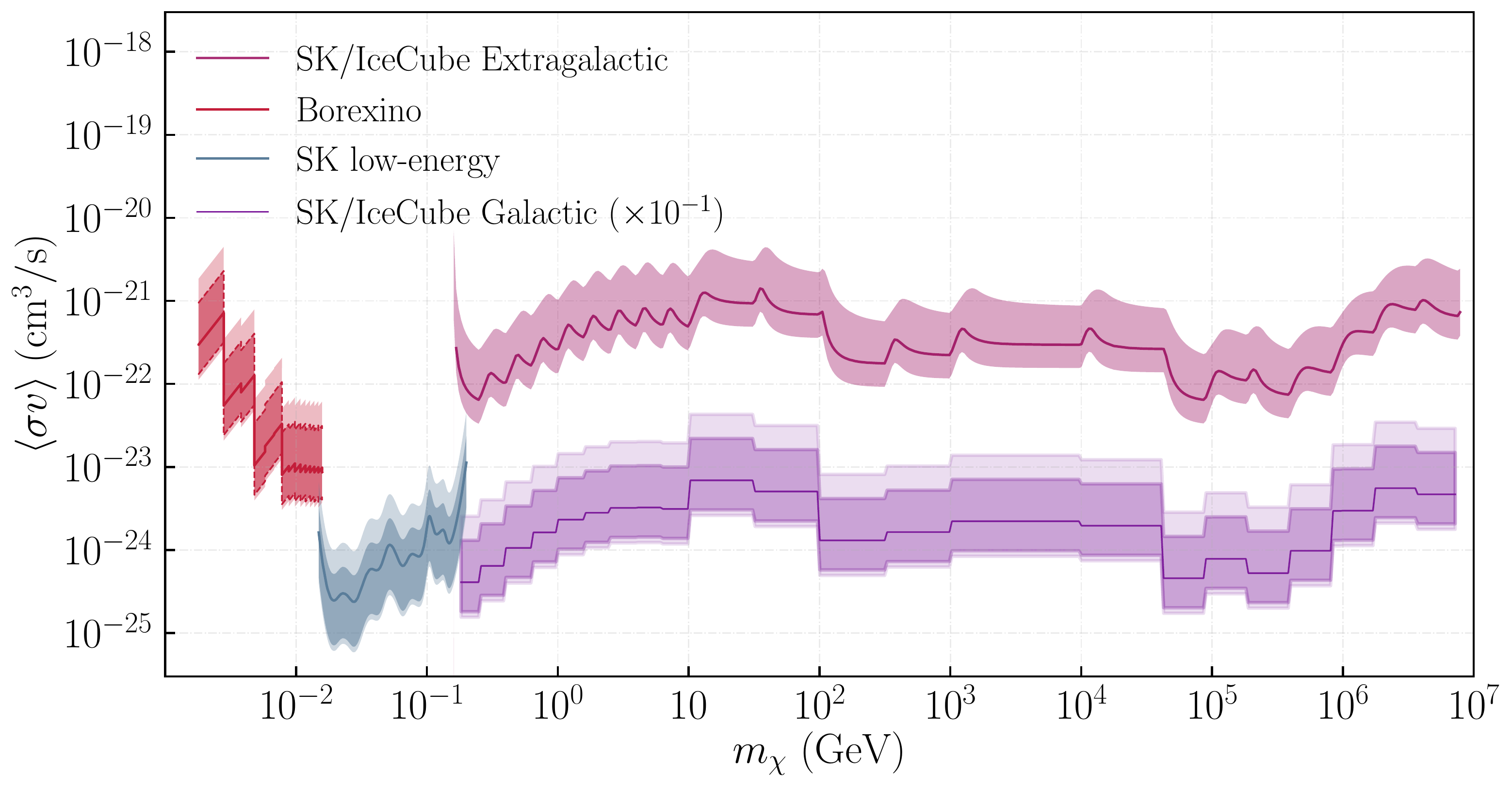}
    \caption{
    \textbf{\textit{Uncertainties on the $s$-wave annihilation cross section for a subset of our results.}}
    Solid lines correspond to the limits discussed in Sec.~\ref{sec:results}.
    For all Galactic limits, namely Borexino (red, leftmost), Super-Kamiokande low-energy (grey, scond region from left), Super-Kamiokande and IceCube (lower, rightmost), the 68\% (dark bands) and 95\% (light bands) uncertainties arise from the allowed variation on the dark matter distribution in the Milky Way, assuming a generalized NFW profile.
    The width of the uncertainty band for the extragalactic limits (upper, rightmost), obtained by comparing to the unfolded neutrino flux from IceCube and Super-Kamiokande, is dominated by the choice of the minimum halo mass, $M_{min}$, although it includes the uncertainty in the choice of HMF $dn/dM$, see Fig.~\ref{fig:hmfparam}.
    For our nominal choice of HMF, we choose the value of $M_{min}$ that yields the weakest constraint.
    }
    \label{fig:sigma_uncert}
\end{figure*}

\textit{\textbf{Milky Way halo shape parameters:}} To quantify the effect of the uncertainty on the MW halo shape parameters, we use the code provided by the authors of \cite{Benito:2019ngh}, which computes the log-likelihood as a function of halo shape parameters \{$\rho_0,r_s,R_0,\gamma$\}, given observed stellar kinematics data.
We profile over the 4 degrees of freedom, modifying the code to account for GRAVITY measurements of $R_0$, and obtain 68\% and 95\% C.L. ranges on the $J$-factors which we propagate to a range on $\sv$ for the Borexino, SK, and IceCube analyses. These are shown as dark and light bands, respectively, in Fig.~\ref{fig:sigma_uncert}.

\begin{figure}[ht]
    \centering
    \includegraphics[width=1\columnwidth]{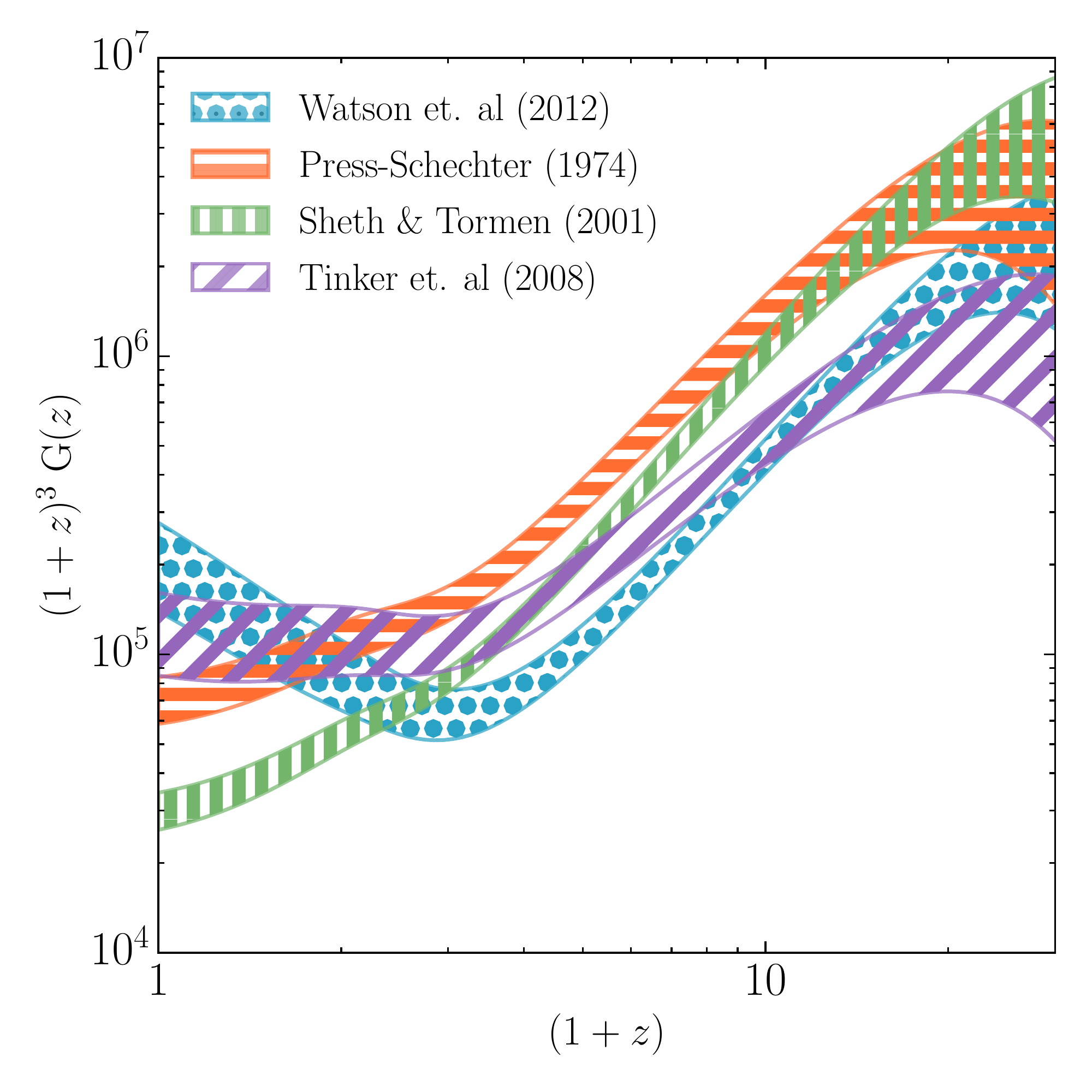}
    \caption{
    \textbf{\textit{The halo boost factor $G(z)$ as a function of redshift for several parametrizations of the HMF $dn/dM$.}}
    Our extragalactic constraints use Watson~\textit{et. al}~\cite{2013MNRAS.433.1230W}. The bands represent varying choices of minimum halo mass, from $10^{-3}$ to $10^{-9}$ solar masses. Fig.~\ref{fig:sigma_uncert} shows the effect of choosing a different parametrization on the limits.
    }
    \label{fig:hmfparam}
\end{figure}

\textit{\textbf{Halo Mass Function uncertainties:}} The largest contributions to uncertainties in the cosmological limits come from 1) the choice of HMF parametrization, and 2) the choice of minimum halo mass, $M_{min}$.
In our analyses we have employed the simulation-driven HMF fit by Watson et al.~\cite{2013MNRAS.433.1230W}.
Fig.~\ref{fig:hmfparam} shows the boost factor $G(z)$ defined in Eq.~\eqref{eq:overdensity}, for four different parametrizations from the literature: the analytic Press \& Schechter formalism~\cite{1974ApJ...187..425P,1991ApJ...379..440B}, Sheth \& Tormen~\cite{Sheth:1999mn,Sheth:1999su}, and Tinker~\cite{Tinker:2008ff}.
The width of the bands comes from varying the minimum halo mass from $10^{-3}$ to $10^{-9} \, M_\odot$. The band labeled ``Extragalactic'' in Fig.~\ref{fig:sigma_uncert} shows how this range propagates through to the cross section constraints.
Since there is no way of statistically quantifying the error on the HMF and minimum halo mass, we choose the most conservative scenario $M_{min} =10^{-3} \, M_\odot$ for our choice of HMF, corresponding to the solid magenta line in Fig.~\ref{fig:sigma_uncert}.

\section{Discussion \& Conclusions\label{sec:conclusion}}

We have presented a comprehensive set of limits on dark matter annihilation directly to neutrino-antineutrino pairs, for a DM mass range from $10^{-3}$~GeV to $10^{12}$~GeV.
Remarkably, there exists uninterrupted coverage of this entire range by the multitude of neutrino detectors that have been in operation over the past decade.
The strongest limits unsurprisingly come from dedicated analyses that include direction and energy information, such as those performed by Super-Kamiokande~\cite{Frankiewicz:2015zma,FRANKIEWICZ:2018lsh}, IceCube~\cite{Aartsen:2016pfc}, and ANTARES~\cite{Adrian-Martinez:2016fdl}.
Unfortunately, such analyses become difficult to accurately recast, as the event information and detector effective area and response are not typically made publicly available. 

Because the DM density is a fixed constraint, the annihilation rate to neutrinos scales as $m_\chi^{-2}$.
A surprising feature of the constraints we have presented here is that they remain approximately flat, rising only two orders of magnitude from $\sv \lesssim 10^{-24}$ cm$^{3}$ s$^{-1}$ to $10^{-22}$ cm$^{3}$~s$^{-1}$ across 9 decades in energy.
Above this range, sensitivity drops off with $\sim m_\chi^2$ since the neutrino cross section only grows logarithmically in this regime.
We attribute the flattening to two main features, which highlight the unique promise of neutrino astronomy: 1) the neutrino-nucleus cross section, which determines the detection efficiency, grows strongly with center-of-mass energy till approximately $E_\nu = 10^6~{\rm GeV}$; and 2) neutrino detectors built for high-energy observations must necessarily be larger, to compensate for the lower expected flux from extragalactic sources, and the larger size of the detectable Cherenkov cascades caused by neutrino interactions.
At energies above $\sim 10^{10}$~GeV, neutrinos become the only probe of high-energy extragalactic processes.

For $s$-channel annihilation, next-generation experiments will finally venture below the expected thermal relic abundance for 10~MeV masses.
In fact, our analysis of the recent SK phase-IV data~\cite{WanLinyan:2018} is within a factor of a few from the relic abundance expected value.
Similarly, with the realization of a cubic kilometer detector in the Northern Hemisphere, the sensitivity in the TeV energy range gets close to the thermal relic expectations.
Beyond the expected thermal relic cross section there are some intriguing hints for dark matter that could be tested with neutrinos, here we mention a few.

The EDGES collaboration recently reported an abnormally low-temperature absorption feature in the 21~cm global spectrum at a redshift of $z \sim 17$~\cite{Bowman:2018yin} though the interpretation of this result has been questioned by a number of studies \cite[e.g.][]{Bradley:2018eev}.
If the observation does hold up to scrutiny and replication, it would be an indication of physics beyond the standard cosmological model.
A suggested explanation is excess gas cooling by millicharged dark matter~\cite{Barkana:2018lgd,Munoz:2018pzp,Klop:2018ltd}, see also~\cite{Berlin:2018sjs}.
In such scenarios, a neutrino line is expected in the 10~MeV range~\cite{Klop:2018ltd}.
This model requires 2$\%$ of the DM to annihilate to muon and tau neutrinos, with a cross section around $ 10^{-25}$ cm$^{3}$s$^{-1}$.
As indicated in Fig.~\ref{fig:Indirect}, this parameter space is rapidly closing. 

\citet{Goodenough:2009gk} noted an excess of gamma-rays seen by the space-borne Fermi-LAT instrument in the direction of the Galactic center in an energy range from 3-10~GeV.
Despite considerable debate, this signal remains consistent with what is expected from DM annihilation~\cite{Leane:2019uhc}, \textit{e.g.} it can be well explained by dark matter annihilation into $b\bar b$ with a mass of $\sim 30~{\rm GeV}$ and an annihilation cross section of the order $10^{-26}$ cm$^{3}$s$^{-1}$~\cite{Hooper:2010mq,Daylan:2014rsa,Calore:2014xka}.
Recent analyses of the AMS-02 cosmic-ray data~\cite{Aguilar:2016kjl} have found hints of an excess in cosmic ray antiprotons, that can also be explained by $\sim 30$~GeV WIMPs annihilating to $W^+W^-$ or $b$ quark pairs with a very similar cross section~\cite{Cuoco:2016eej}.
The detection of a complementary neutrino signal to what is seen in the GC would be a powerful indication of new physics processes at work. Caution is warranted, as the antiproton excess could well be attributed to systematic uncertainties in cosmic ray propagation~\cite{Boudaud:2019efq} or a combination of propagation uncertainties, nuclear cross section uncertainties, and correlations in instrumental systematics
~\cite{Heisig:2020nse}.

Additionally, growing statistics for different channels for observation of high-energy neutrinos in IceCube~\cite{Aartsen:2016xlq,Schneider:2019ayi} hints towards a more complex spectral scenario and possible features in the flux of cosmic neutrinos.
Analysis of the contained neutrino events at lower energies ($\sim 10$~TeV) has revealed a flux that is an order of magnitude higher than the flux at PeV energies~\cite{Aartsen:2014muf}.
This is usually referred to as the ``low-energy excess" in IceCube data.
The origin of these neutrinos are thought to be different from the bulk of neutrino emission at PeV energies, see~\cite{Murase:2015xka} for more discussion.
Interestingly, models assuming DM annihilation (or decay) into high-energy neutrinos have been proposed to describe the low-energy excess~\cite{Chianese:2017nwe,Bhattacharya:2019ucd}, see also~\cite{Sui:2018bbh}, and they show a slight preference for a potential component from TeV dark matter. However, such interpretation could be in tension with gamma-ray observations~\cite{Chianese:2018ijk}.
At the moment, it is clear that elucidating the origin of the high-energy neutrino excess will require correlated observations with gamma-rays and novel analysis techniques, see \textit{e.g.}~\cite{Dekker:2019gpe}.

The ANITA balloon-borne experiment has recently reported on two events originating from 30$^\circ$ or more below the horizon~\cite{Gorham:2016zah,Gorham:2018ydl}, with energies in excess of 500~PeV.
This is unexpected, as the Earth should be opaque to neutrinos at these energies.
These are not consistent with either a diffuse primary neutrino flux, or a point source hypothesis, as the secondary interaction products would have been observed at IceCube~\cite{Romero-Wolf:2018zxt,Safa:2019ege, Aartsen:2020vir}.
Systematic effects regarding irregularities in the Antarctic surface ice have been proposed~\cite{Shoemaker:2019xlt}.
However, dark matter which decays~\cite{Cline:2019snp,Hooper:2019ytr} or annihilates~\cite{Esmaili:2019pcy} to neutrinos or boosted DM could also explain such a signal, though more data are still required to test such hypotheses~\cite{Anchordoqui:2019utb,Dudas:2020sbq}.

We hope for further surprises and point out the great room for improvement with dedicated analyses; \textit{e.g.} our DUNE and HK estimations do not yet use directional information.
Likewise, high-energy neutrino observatories are expected to improve their angular and energy resolutions in the next generation and a combination of their data sets would improve over our projected sensitivities.

The annihilation of dark matter to neutrino pairs is the most invisible channel: the constraints that we have provided here are thus closing the window on dark matter annihilation into standard model products, and are thus rapidly narrowing down the available parameter space where WIMP-like dark matter may still be hiding. 

\FloatBarrier

\section*{Acknowledgements}
We would like to thank John Beacom, Mauricio Bustamante, Claire Gu\'epin, Francis Halzen, Julian Heeck, Matheus Hostert, Teppei Katori,  Gordan Krnjaic, Elisa Resconi, Andr\'es Romero-Wolf, Carsten Rott, and Sergio Palomares-Ruiz for useful discussions. We thank the anonymous referees for excellent suggestions and comments.
We are grateful to have had the chance to work with our friend and colleague AOC, and wish him the best success in future endeavours. 
CAA and AD are supported by NSF grant PHY-1912764.
AK acknowledges the IGC Postdoctoral Award. IS is supported by NSF funding support PLR-1600823 and OPP-1600823.
ACV is supported by the Arthur B. McDonald Canadian Astroparticle Physics Research Institute, with equipment funded by the Canada Foundation for Innovation and the Ontario Ministry of Economic Development, Job Creation and Trade (MEDJCT).
Research at Perimeter Institute is supported by the Government of Canada through the Department of Innovation, Science, and Economic Development, and by the Province of Ontario through MEDJCT.

\bibliographystyle{apsrmp4-1}
\bibliography{nudm}
\end{document}